# A novel vortex-assisted generation system for hydrokinetic energy harvesting from slow water currents

**Ulugbek Azimov\*, Jack Callaghan, Iman Frozanpoor, Callum Hulsmeier, Christopher Kwok, Matthew Nixon**

Mechanical and Construction Engineering Department, Northumbria University, Newcastle upon Tyne, NE1 8ST, United Kingdom
\*Corresponding Author: ulugbek.azimov@northumbria.ac.uk

**Abstract**
A novel system has been developed that harnesses the phenomena of vortex-induced vibrations (VIV) from a slow current (<0.5 m/s) of water to generate renewable hydrokinetic energy. It utilizes a single degree-of-freedom pivoting cylinder mechanism coupled with an electromagnetic induction generator. As a result of observation and concept development, the final prototype includes a stationary cylindrical shedder upstream of the oscillator. The system is referred to as 'Vortex Assisted Generation' (V.A.G.) throughout the report. Given the novelty of the system, an extensive investigation has been conducted to identify key parameters and functional relationships between system variables, regarding their effect on output voltage, frequency, and power. A range of flow velocities have been established that instigate system lock-in, where the cylinder oscillates at high amplitude and frequency. For the tested prototype up to 37% of system extraction efficiency has been achieved in the lab conditions.

**Keywords:** Vortex-induced-vibrations; Hydrokinetic energy; Slow current flow; Oscillating bluff body; Fluid-structure interaction.

## 1. Introduction

With the increasing energy demand, there is a need for a significant shift to using clean and renewable sources of energy. Whilst tidal power is a more reliable energy source than wind or solar power, the development of tidal power plants is hindered by unpredictable wave intensities, high cost and the scarcity of ideal vacated locations. Recently a new paradigm to extract energy from currents has been developed which is based on vortex-induced motions. In this paradigm, the energy of vortices is recovered instead of providing flow with extra energy artificially. Vortex shedding, due to which vortices are generated and detached from the body, changes the local pressure distribution around the body. This local change in pressure distribution induces the motion on the body. When a bluff body is subject to a relatively mobile fluid, or the body itself is relatively mobile within a fluid, if the flow conditions are appropriate and the bluff body is in an appropriate degree of freedom configuration, separation of the boundary layer begins to take place on the surface of the body. This causes a transient pressure differential around the edge, which results in oscillation of the body [1]. This phenomenon is named vortex-induced vibration (VIV). The separation of the boundary layer from the surface causes vortex formation in the wake of the body, known as a von Karman vortex street. VIV, capable of utilizing hydrokinetic energy from slow moving currents, provides an alternative solution to clean energy harvesting. This cost-effective alternative of harnessing of aquatic energy from relatively low flow rates presents a viable solution providing a reliable output coupled with high efficiencies.

Real world examples of VIV include riser tubes for oil extraction that are influenced by the ocean current, or the pipes in a crossflow heat exchanger that are influenced by the transfer media flow [1]. The conditions for VIV are mainly determined by the Reynold number of the flow around the bluff body, which is determined by the flow speed, bluff body geometry and flow viscosity. To optimally maximize the amplitude of VIV the vortex shedding frequency needs to be as close to the natural frequency of the system as possible.

To generate usable power from VIV, mechanical energy must be converted into electrical. A most significant work to harness energy from VIV phenomenon was patented by Bernitsas et al. [2]. Their energy converter is called VIVACE (Vortex Induced Vibrations for Aquatic Clean Energy) that uses a passive circular cylinder with upward-downward motion, induced by vortex shedding. This system utilizes a cylindrical body positioned horizontally perpendicular to the flow, which oscillates reciprocally and elastically by mounting to linear springs. It is connected to a power take-off system via a transmission mechanism, which combines a gear-belt system and a low rpm electrical generator. VIVACE is scalable and can extract energy from currents of velocities 0.25 to 2.5 m/s. For VIVACE system the water must be deep enough to permit the safe passage of boats, and to avoid interaction between the wake of the converter and both the surface and the riverbed or seafloor. The maximum peak efficiency for this system was reported to be around 30.8% [3]. Another method for generating electricity using VIV is the Vortex-Induced Piezoelectric Energy Harvester (VIPEH) system [4]. This consists of a lead zirconate titanate (PZT) and copper cantilever with a cylindrical extension. As the cylinder vibrates, the cantilever is put under strain and an electrical current is generated. Electromagnetic (EM) induction can also be used for the conversion of mechanical to electrical power from VIV. This proposed method utilizes a tubular linear interior permanent magnet generator to minimize mechanical components [5]. It is suggested that this method of generation is more robust and modular than VIVACE. Also another system that utilizes EM induction for power generation involves a flexible diaphragm that deforms by the force of the vortices and moves a magnet into the path of a coil [6].

Over the past two decades, researchers have studied the downstream wake characteristics using visualization techniques. For example, Williamson and Roshko [7] investigated the wake flow induced by a forced-vibrating cylinder under different vibrating modes. They found that the flow pattern can be categorized into two modes, namely the 2S and 2P. Brika and Laneville [8] observed the foregoing two wake modes induced by a freely vibrating long flexible pipe by using flow visualization techniques. Govardhan and Williamson [9] used PIV to measure the velocity field of different vibrating modes of cylinders. They observed that for low reduced velocities the pattern of the wake vortices typically appeared as the 2S mode. Shang et al. [10] discovered the occurrence of the P+S mode at very high reduced velocities, whereas as Zhao et al. [11] found the existence of the 2S mode in the upper branch. Wang et al. [12] also used PIV to investigate the effect of a planar boundary on vortex formation.

A great number of experiments on VIV deal with a rigid cylinder submerged in water and subjected to a steady flow. These types of experiments can be divided into three distinct groups: (1) stationary experiments, in which the cylinder is fixed and acting forces are measured [13]; (2) forced vibration experiments, in which an immersed cylinder is forced to move with a given amplitude and frequency, and the fluid forces are registered [14]; and (3) free vibration experiments, in which a cylinder can vibrate freely in the flow, and its motion is recorded [9, 14, 15, 16, 17, 18, 19]. During the free vibration experiments, the amplitudes and frequencies of the cylinder's motion are recorded and subsequently used to tune empirical models that simulate the fluid-structure interaction, such as the wake-oscillator model [20, 21]. The important parameters that affect the energy extraction by VIV are the mass ratio, the mechanical damping, the reduced velocity and the Reynolds number [22]. Lee and Bernitsas [23] concluded by experiments that the damping has a strong influence in the maximum efficiency attainable. To consider the effects of different parameters on VIV performance, Barrero-Gil et al. [22] have performed a parametric study by mathematical modelling on a cylinder which has sinusoidal oscillation steadily by amplitude and frequency. They have formulated the efficiency in terms of normalized amplitude, normalized velocity and normalized frequency. Hsieh et al. [24] investigated the flow characteristics around a circular cylinder undergoing VIV. They studied the vortex structure and turbulence statistics in the wake of a freely-vibrating cylinder using visualization techniques and a high resolution PIV. They found that the vibration of the cylinder results in the formation of oblique upward and downward jets. They also found that the size of the affected wake zone in proximity of the bluff body is much wider for the vibrating case compared to the stationary case. Chang et al. [25] has proposed a creative method called passive turbulence control to enhance the power harnessing from VIV of circular cylinder.

Ding et al. [26] studied the flow induced motion and energy harvesting of bluff bodies with different cross sections. They found that the flow induced motion responses of passive turbulence controlled-cylinder and quasi-trapezoid body are stronger than the square cylinder and triangular prism. They confirmed that energy can be obviously harvested by the different cross-section cylinders when Reynolds number exceeds 30,000.

Barbosa et al. [27] investigated VIV of a freely vibrating cylinder near a plane boundary. It was observed that inside the lock-in region, the amplitude of oscillation is not affected by the plane boundary if the gap between the cylinder and the boundary is larger than two diameters. For gaps between 0.75 and 2 diameters, the amplitude of oscillations tend to decrease, but the oscillations remained symmetric with respect to the equilibrium position, and no impact with the boundary is observed. For gaps smaller than 0.75 diameters, the cylinder will impact the boundary, resulting in a non-symmetric oscillation. There are a number of publications on experiments aimed at studying the effect of a boundary on the fluid-structure interaction forces for the case of stationary cylinders [28, 29, 30, 31]. Important conclusions of these studies are that vortex shedding is suppressed for gaps smaller than one third of the cylinder's diameter, and that there is a mean cross-flow force that pushes the cylinder away from the boundary. Nevertheless, an important conclusion is that also for small gaps (even smaller than 0.3D), VIV can occur. This conclusion implies that the motion of the cylinder stimulates vortex shedding, a phenomenon that is suppressed in cases where the cylinder is fixed. Based on the experimental results, an extended wake-oscillator model [20, 21] capable of representing the boundary effects was proposed.

Kim and Bernitsas [32] predicted the performance of horizontal hydrokinetic energy converter using multiple-cylinder synergy in flow induced motion. They found that high power-to-volume density and high power conversion efficiency was achieved by the multi-cylinder converter even in river or ocean currents as slow as 1.0 m/s, where marine turbines don't work efficiently. Sun et al. [33] studied the hydrokinetic energy conversion by two rough tandem-cylinders in flow induced motion. They have shown that in the VIV range the improvement due to the presence of the downstream cylinder is minimal and not worth the cost of designing a more complex converter. The two-cylinder converter can harness energy from flows as slow as 0.4 m/s at low harness damping and lower stiffness. To harness the power they introduced the damping to the system. This resulted in reduction of the oscillator amplitude in both VIV and galloping ranges. Damping delayed the onset of galloping.

Zhang et al. [34] studied vortex-induced vibration for a single-degree of freedom isolated circular cylinder under the wake interference of an oscillating airfoil. They found that for vortex induced vibration case at the different wakes of the oscillating airfoil, the vortices shedded from the oscillating airfoil boundary layers will be mixed with the vortices created by the vibrating cylinder and generate concentrated vortices in both the drag and thrust producing cases.

Although a great number of experiments and numerical studies have been carried out on VIV, they all have looked at horizontally positioned and fully submerged bluff bodies. As far as the VIV application is concerned, these configurations are well suited to harvest energy from deep waters via tidal and wave implications. However, they are not quite suitable to be used in slow and shallow waters, in which category most of the small rivers and canals fall. The motivation for the experiments and consequently the design of a novel vortex-assisted generator system was the scarcity of reports on freely vibrating vertical cylinders near a plane wall and the shedder. The concept and design of the vortex-assisted generation system proposed in this paper allows harvesting the hydrokinetic energy in slow and shallow waters. In this contribution, the results obtained during the free vibration experiments on the vertical pivoted cylinder for assessing the influence of vortex shedder on a system's efficiency has been reported. It will consider the constraints involved in designing a single degree-of-freedom system incorporating a novel arrangement to convert mechanical motion into a useful electrical output. This work will not address the widely studied lock-in phenomenon; instead the focus will be on the effect of the flow velocity, the bluff body submersion level and the frequency on the voltage and system extraction efficiency, which has important implications in energy conversion. The experiments will cover a range of flow velocities, different shedder distances and submersion levels to better understand the vortex formation characteristics, the pressure distribution around the cylinder and its effect on the coefficient of lift. The obtained extensive dataset provides further insight in the behavior of structures with a vertical pivoted cylinder, thus allowing for more accurate predictions of the dynamic behavior of such systems.

## 2. Theoretical background

Vortex shedding is the phenomenon where eddy currents are produced either side of a bluff body, due to the flow speed of the fluid being higher on the outside of the wake. The phase and hence production of these currents depends on the geometries of the body. The presence of eddies creates an alternating pressure distribution, which effectively moves the body in the axis normal to the flow direction. As the swirling vortices flow downstream they create a repeated pattern in the shape of a sinusoidal wave, known as a von Karman vortex street [35]. The vortex shedder can harness the energy produced from the currents, if the body is free to move in the direction normal to the flow, the vortices can cause it to oscillate. This is known as Vortex-Induced Vibration (VIV). The magnitude of the vortices and thus the size and speed of the oscillations are determined by the bluff body geometry and fluid properties. If the contributing factors provide an ideal regime for vortex shedding, then a design can be created with the aim to efficiently convert the hydrokinetic energy into a linear motion [22]. A number of contributing variables determine the wake regime of a bluff body and flow rate. From this, geometries can be established for slow currents, below 1 m/s. The ideal regime for VIV is defined by the Reynolds number. The wake, which is defined as the stream of vortices, produced by the vortex shedder is represented by a range of Reynolds numbers, which is directly related to the fluid and bluff body properties. This was first proposed by Lienhard [36]. He showed that at very low Reynolds number the streamlines of the resulting flow is perfectly symmetric as expected. However as the Reynolds number increases the flow around the cylinder becomes asymmetric and the separation of boundary layers begins to occur, eventually leading to the production of eddies which break away from the cylinder. The wake can be altered whilst in a constant fluid flow; this means the magnitude of the wake, depending on the Reynolds number can be independent of the fluid supply velocity. The same figure shows how the stream of vortices changes with the Reynolds number. This data can be used to create a system that produces a desired wake (i.e. a laminar vortex street will be produced at Reynolds numbers between 10 and 150.

As a dimensionless parameter, the Strouhal number represents the ratio of the inertial forces due to the unsteadiness of the flow. It is directly related to the Reynolds number.

$$S_t = \frac{f_s d}{V} \tag{1}$$

This parameter should be maintained at 0.2 [1], in order to achieve a fully developed von Karman Vortex Street. It relates the frequency of vortex shedding $f_s$ to the velocity of the flow $V$ and a characteristic dimension of the bluff body $d$, depending on its shape.

At low Reynolds numbers ($10^2$ to $10^5$), or subcritical flows, the Strouhal number should be maintained at 0.2 in order to achieve vortex shedding. With a more turbulent flow, the Strouhal number has a larger range, as shown in [1]. Strouhal instability occurs when the vortex shedding frequency becomes close to the natural frequency of vibration of the bluff body. Defined as the lock-in phase, the largest magnitudes of vibration are held within this phase [1, 37]. The shedding frequency can be found from the Strouhal number, which is useful when trying to define the lock-in phase of a system/structure. If the vortex shedder is free to move in the direction normal to the flow it is possible that the vortices produced will cause the body to oscillate from side to side, in sync with the alternating pressure distribution.

### *Shedding Frequency & Lock-In*

Lock-in phase, sometimes referred as synchronization, is often perceived as the regime where the frequency of vortex shedding is approximately equal to the natural frequency of the structure/system. This has been known to produce the maximum amplitudes of oscillations, $f_s \approx f_n$. Synchronization is said to be achieved when the two frequencies are within 20% of each other [38]. The shedding patterns in the wake of an oscillating cylinder exist for multiple ranges of frequencies and amplitudes [39]. Along with these modes of shedding, the force on the cylinder and hence the lift varies.

Williamson and Roshko [7] illustrated the numerous modes that could be utilized in vortex induced vibrations; the most dominant modes include 2P, 2S and P+S. Other than the distinctive modes, it is possible that the vortex shedding from a cylinder will produce other shedding patterns. These additional patterns such as C(P+S), C(2S), 2P+2S and C are separated by the critical curve. This curve marks the transition from one mode to another. The two distinct modes are '2S' and '2P'. 2S sheds two single vortices per cycle whilst 2P sheds a pair of vortices either side of the bluff body per cycle following the sinusoidal vortex street. Different modes of vortex shedding are defined by the reduced velocity '$U^*$' and the amplitude response '$A^*$' of the VIV system. Williamson and Roshko [7] showed the transition of the vortex shedding modes in respect to these parameters.

$$U^* = \frac{U}{f_n D} \tag{2}$$

$$A^* = \frac{A}{D} = \frac{1}{(m^* + C_A)\zeta} \left[ \frac{C_Y \sin\varphi}{4\pi^3} \left( \frac{U^*}{f^*} \right)^2 f^* \right] \tag{3}$$

A typical VIV system consists of a bare cylinder restricted to only move freely in either the vertical (VIVACE) [2] or horizontal axis (VMDS) [40]. Normally the system has zero damping effect and the cylinder is able to move freely along the axis. Therefore the oscillations of the body are controlled only by the vortices shed by the shedder. A simple set-up of a cylinder following simple harmonic motion is shown in Fig. 1.

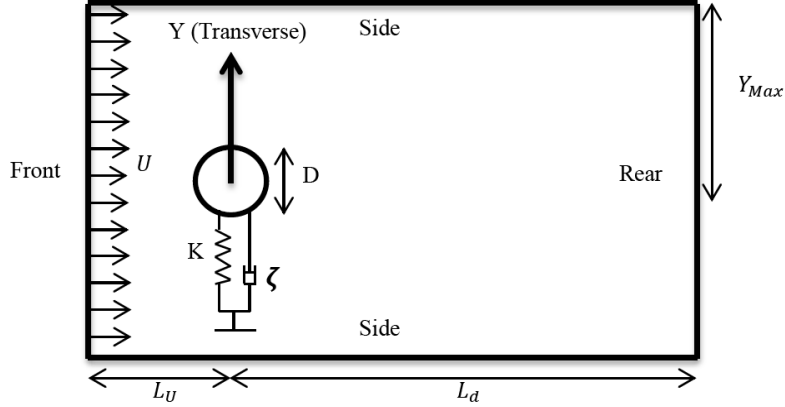

Fig. 1. Schematics of a cylinder harmonic motion system

## *Equations of Motion*

The motion of vortex induced vibrations of the cylindrical body is formulated by considering the set-up to be a single degree of freedom system.

$$m\ddot{y} + c\dot{y} + ky = F_{fluid} \tag{4}$$

Cylinder Displacement: $y(t) = A sin(\omega t)$ (5)

Fluid Force: $F(t) = F_0 sin(\omega t + \varphi)$ (6)

The response amplitude is derived from the equations above, of which the response frequency can also be derived.

$$f^* = \left( \frac{m^* + C_A}{m^* + C_{EA}} \right)^{\frac{1}{2}} \tag{7}$$

Here, '$C_A$' is the potential added mass coefficient, and '$C_{EA}$' is the 'effective' added mass coefficient, which includes the apparent effect due to the total transverse fluid force in-phase with the body's acceleration ($C_y cos(\phi)$).

$$C_{EA} = \frac{1}{2\pi^3} \frac{C_y cos(\varphi)}{A^*} \left( \frac{U^*}{f^*} \right)^2 \tag{8}$$

It is worth noting that the response amplitude is proportional to the transverse force component that is in phase with the body's velocity and for small masses and damping effects; the precise value of the phase angle '$\phi$' has a large effect on the amplitude.

## *Critical Mass*

The synchronisation regime becomes infinitely large when the mass ratio decreases to a critical value of 0.54±0.2. This value depends on the vortex shedders cross sectional shape. For the free vibration of a cylinder, with negligible mass damping, lower amplitudes determine the upper end of the lock-in regime. The amplitudes give a constant oscillating frequency which is proportional to the reduction in mass.

$$f^*_{Lower} = \left( \frac{m^* + 1}{m^* - 0.54} \right) \tag{9}$$

This provides a simple method of calculation of the highest frequency attainable of a VIV system in the lock-in phase.

As the system oscillates the former (and coil) move relative to the magnets creating a change in magnetic flux density relative to the coil inducing an EMF (electromotive force) within the coil. This process can be described by Faraday's Law of Electromagnetic Induction:

$$\varepsilon mf = -N\frac{d\Phi}{dt} \tag{10}$$

Where $N$ is the number of turns on the coil and $\Phi = BA$, where $B$ is the magnetic flux density (Tesla) and $A$ is the cross sectional area of the coil.

As the machine has a fixed area coil and multiple magnetic poles, the equation can be rewritten as:

$$\varepsilon mf = -N \cdot A \cdot P\frac{dB}{dt} \tag{11}$$

where, P is the number of magnetic pole pairs.

As the generator consists of a coil the inherent impedance consists of both resistance and inductive reactance where inductive reactance is expressed as:

$$X_L = 2\pi \cdot f \cdot L \tag{12}$$

where, $L$ is the coil inductance and $f$ is the generation frequency.

Due to the inductive nature of the generator the real output power ($P$, expressed in Watts) is equal to the apparent power ($S$, expressed in VoltAmps) multiplied by the cosine of the power factor.

Apparent power is expressed as:

$$S = I \cdot V_{RMS} = \frac{V_{RMS}^2}{Z} = \frac{V_{RMS}^2}{\sqrt{R^2 + X_L^2}} \tag{13}$$

And power factor is:

$$p.f. = \tan^{-1}\left(\frac{X_L}{R}\right) \tag{14}$$

Therefore the real power output can be expressed as:

$$P = \frac{V_{RMS}^2}{\sqrt{R^2 + X_L^2}}\cos\left(\tan^{-1}\left(\frac{X_L}{R}\right)\right) \tag{15}$$

## 3. Concept Development

Six initial designs were considered as shown in Fig. 2. The green arrows show the respective motion of the bluff bodies where the bridges are designed to sit perpendicularly to the flow in the channel. A cylinder was used as the chosen bluff body geometry as it is the most used in successful applications. 25 mm was the selected diameter as this is the maximum effective size that can be used in the available channel.

In parallel to mechanical design, electrical power generation has been considered. In order to convert the kinetic energy within the flow of water into electrical power, a generator is required. Three systems of electrical generation have been considered: piezoelectric, electromagnetic induction and electrostatic induction. Of the three options electromagnetic induction has been selected as the most appropriate due to the potential scalability and simplicity of design. To minimize mechanical losses between the pivoted cylinder and the generator a direct driven linear machine has been designed. The linear machine consists of a stator rod on to which neodymium magnets are mounted, and a former (mover) with an enamelled copper coil winding which is attached to the pivoted cylinder.

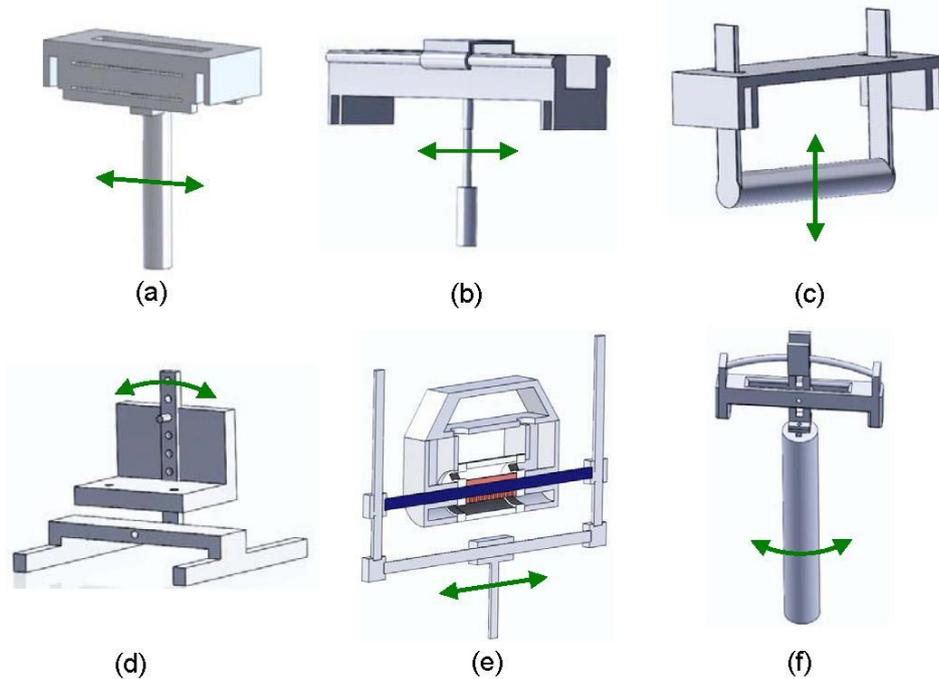

Fig. 2. Initial concepts. (a) Linear design consists of 4 simple bearings and a static bluff body, (b) Linear design consists of a rail track with a static bluff body, (c) Linear design which utilizes the buoyancy and gravity, based on VIVACE concept, (d) Simple pendulum design with a dynamic bluff body which can change its vertical height, (e) An air bearing design which removes all of the mechanical friction in the design. The bluff body can move freely in two directions as well as locking the body to only move freely in one direction, (f) Simple pendulum design with a single bearing.

From the initial testing it was seen that there were issues with the alignment and rigidity of the generator relative to the pivoted arm. To combat this, a new frame was developed that consisted of two sandwich plates of clear PVC that clamped all of the static components in a uniform plane. Also, the moving parts were now all nut clamped to the main rod that ran the length of the swinging arm. It was noticed that placing another static bluff body directly upstream of the pivoted cylinder, oscillations occurred at a higher range of flow velocities. From this observation another test plan was made and the current system was retrofitted to incorporate another static body or 'shedder' as shown in Fig. 3.

All testing was to be completed on the same system to ensure continuity of experimentation, which allowed a new concept to be generated in the meantime. A build-up of oxidation was noticed along the main arm rod near to the submergence side. This, along with the decision to remove as many ferrous parts from the generator area (to minimize magnetic attraction), led to the idea that the main rod would be made out of PVC. This change also increased robustness and improved alignment of the system, as all fits are square cross-sections instead of round.

VIV energy production occurred easier in the wake of the static bluff body with this the system did not need to be taken to an optimum region where lock-in occurred. Instead what was found was that the generator oscillated from a given flow velocity, after which the frequency and output voltage increased.

To clearly identify the energy harvester throughout the project the name 'Vortex Assisted Generation' (V.A.G.) was given, and a logo was created which allows product recognition if the direction of patent/practical application is ever pursued. This can be seen in Fig. 3. The sinusoidal wave represents the movement of the bluff body as a driven-damped pivoted mass, and the spiral represents the vortices that are the driving phenomenon behind the whole concept.

Prototype 1 had an adjustable frame to allow for height adjustment when placed on the channel; this caused alignment issues hence Prototype 2 consisted of a rigid frame with a set height, allowing the generator to be correctly aligned. Prototype 1 utilizes VIV directly as the oscillator causes the vortices to be formed, however Prototype 2 includes an additional bluff body (shedder) to induce the vibrations on the oscillator further down steam. To accommodate the shedder into Prototype 2 the bridge design incorporates an additional component that allows for the space between the bluff bodies to be adjusted. Less material is used in Prototype 2; most parts are able to be produced through rapid manufacturing techniques and consist of lightweight and cheap polymers.

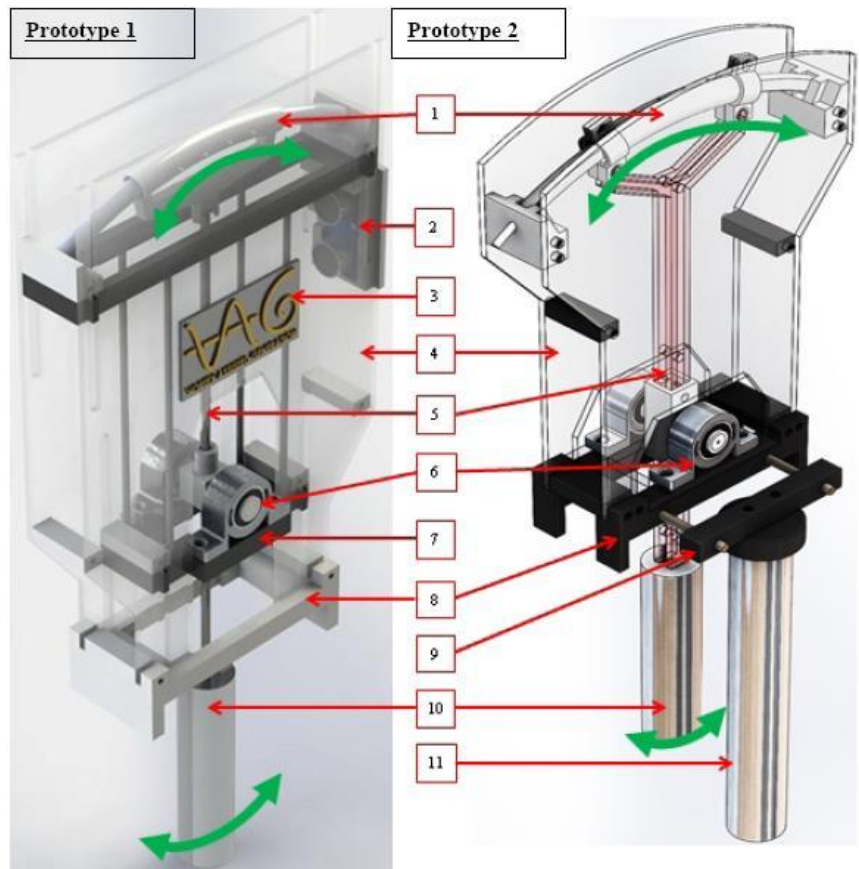

Fig. 3. Vortex-Assisted Generation (V.A.G) system prototypes. 1-Curvilinear generator, 2-Ultrasonic sensor mount, 3-Logo, 4-Transparent body, 5-Pivot arm, 6-Pillow bearing, 7-Height stage, 8-Bridge, 9-Adjustable shedder mount, 10-Oscillator bluff body, 11-Shedder.

The capacity of oxidation of the pivot arm in Prototype 1 is removed in Prototype 2 by changing the material to a polymer. This is especially an important consideration for salt water applications. Alignment in Prototype 2 of the curvilinear stator ring of the generator was further improved by utilizing a keyway in the spacer and internal stator ring design, and improving the way in which this attached to the frame structure. Both designs consist of a frame structure that allow for external effects from environmental conditions such as wind to be minimized by shielding the generator. Prototype 1 was prone to issues of alignment caused by the threaded pivot arm bar which would rotate slightly in application and constantly required to be checked and adjusted in application. Prototype 2 employs a pivot arm made of two acrylic arms that fit into a square pivot point and allows for perfect alignment that remains in the correct position.

## 4. System Power Output & Efficiency

Analysis of the efficiency of the V.A.G. system has been broken down into two parts in order to determine the system's ability to extract power from the flow and convert the extracted power into electrical power. Due to constraints i.e. the length of the flow channel, an accurate value of the input power in the flow is unobtainable as a measurement of flow velocity can only be taken downstream of the V.A.G. without disrupting the input to the system. As such the input power in the flow has been approximated as the sum of the power in the flow within the systems swept area downstream of the system (the extraction loss) and the mechanical power gained by the system. The extraction loss can be expressed as:

$$P_{Loss} = \frac{1}{2} \cdot \rho \cdot A \cdot V^3 \tag{16}$$

where, $\rho$ is the fluid density, $A$ is the swept area and $V$ is the flow velocity.

The swept area is the cross sectional area of fluid across which the pivoted cylinder moves as shown in Fig. 4.

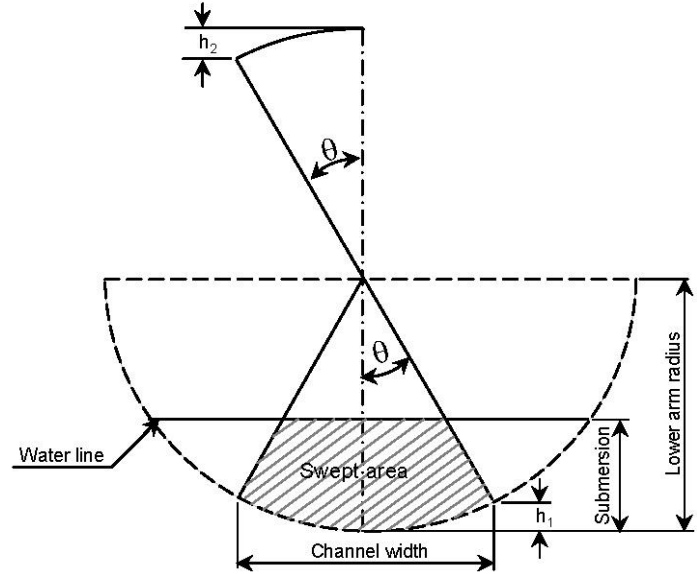

Fig. 4. Swept Area of Flow by Pivoted Cylinder

From Fig. 4 it can be seen that θ is given by:

$$\theta = \sin^{-1}\left(\frac{\left(\dfrac{w}{2}\right)}{R_L}\right)$$
(17)

And the swept area is given by:

$$A = R_L^2 \cdot \theta - (R_L - S)^2 \cdot \tan(\theta)$$
(18)

where, $R_L$ is the lower arm radius, $w$ is the channel width and $S$ is the submersion.

The mechanical power gained by the system is determined as the change in potential energy over time. The system can be considered as a combination of a pendulum and inverted pendulum. From Fig. 4 it can be seen that the change in potential energy within the system can be expressed as:

$$\Delta PE = M_1 \cdot g \cdot \Delta h_1 - M_2 \cdot g \cdot \Delta h_2$$
(19)

where, $M_1$ is the lower arm mass, $M_2$ is the upper arm mass, $\Delta h_1$ is the change in height of the lower arm and $\Delta h_2$ is the change in height of the upper arm.

From this, the mechanical power gained by the system can be given as:

$$P_{Mech} = 2 \cdot \Delta PE \cdot f$$
(20)

where, $f$ is the system oscillation frequency.

Therefore the total power input is approximated as:

$$P_{in} \approx P_{Loss} + P_{Mech}$$
(21)

This neglects any channel losses.

The systems extraction efficiency describes how effective the system is at extracting power from the flow and is given by:

$$\eta_{ext} = \frac{P_{Mech}}{P_{in}}$$
(22)

The system efficiency refers to the system's ability to convert mechanical power to electrical power and is expressed as:

$$\eta_{sys} = \frac{P_{out}}{P_{Mech}}$$
(23)

where, $P_{out}$ is the electrical power output from the system.

## 5. Testing and performance analysis

In order to develop a system of maximum power output and efficiency, a better understanding of the system parameters was required. To achieve this in the absence of theory (due to the novelty of the design), empirical testing was conducted to generate data, which was interpreted to determine how parameters affect energy generation. In order to control the number of experiments and reduce the amount of testing required, variables were controlled. These included:

1. A solid aluminium bluff body was used as the vortex shedder as the weight would mean the force of inertia would overcome the force of buoyancy and the damping effects of the system

2. The diameter of the bluff body was kept constant (25 mm) throughout all experiments. This was due to constraints from the set flow channel dimensions.

3. The upper arm length for the generator was maintained at 160 mm, which reduced the time needed to fabricate parts and test models

Assumptions included what was expected to be seen from empirical testing. This allowed for the creation of a simple but thorough methodology for testing all parameters in order to define their relationships with one another and the flow properties. Assumptions included:

1. Negligible damping due to friction at bearings and stator ring-rotor arm interfaces

2. Flow was fully developed at the 3 m mark of the 5-meter long flow channel

3. Negligible channels losses (Power Upstream = Power Extracted + Power Downstream)

Experiments were conducted using 5-meter long flow channel with controlled flow velocity as shown in Fig. 5. Multi-channel oscilloscope was used for data acquisition. The prototypes shown in Fig. 6 were fabricated to be adjustable in order to examine various values of variables such as lower arm radius, submersion level and flow velocity. The aim of this was to find the optimum settings of the system in various flow rates, all below 1 m/s.

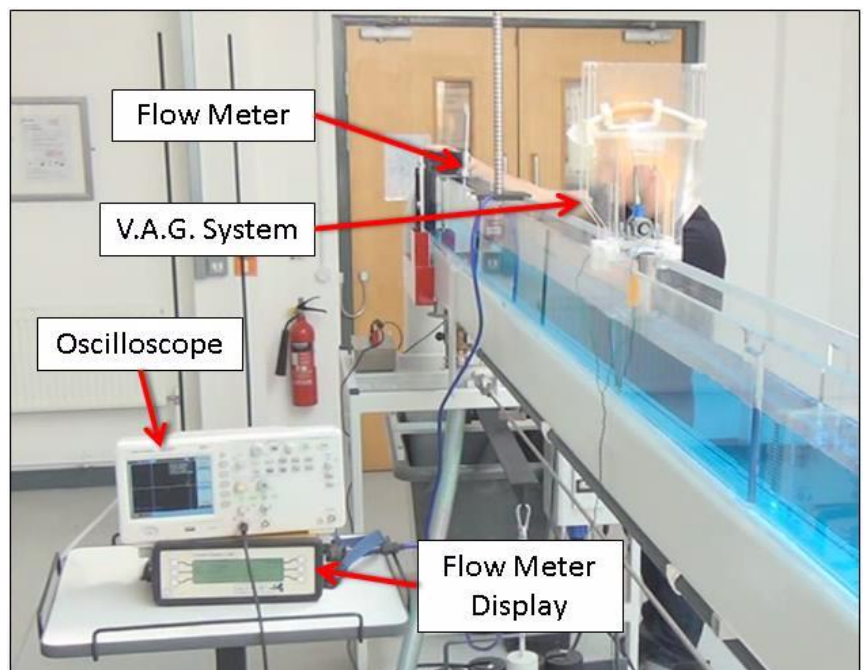

Fig. 5. Experimental setup

Two testing methodologies were provided: the first is to examine the system and corresponding performance when VIV are used to oscillate the system directly; and the second is to examine the performance of the same system when a bluff body is used to generate vortices and cause the system to oscillate in the wake of produced vibrations, VIV indirectly driven system. The material of bluff body was aluminium, and the diameter of the bluff body was 25 mm.

For the first testing we had to determine how lower arm radius, level of submersion, and flow velocity affected the system's performance in terms of power generation on a system that generates vortices and utilizes direct VIV energy generation. Given the experimental set-up and constraints with the flow channel, flow velocities were recorded as an output once lock-in phase was achieved. The transition point to this phase was identified as the point when the cylinder began to oscillate. We ensured the proper alignment of the system, in that the rotor arm coil can oscillate end to end of the stator ring containing the permanent magnets. Given the adjustable nature of the model tested we ensured all nuts are tightened to reduce risk of miss-alignment during operation. The lower arm radius was tested in a range of 180, 190, and 200 mm from the central pivot to the lower end of the bluff body and adjusted to correct radial length. The system was placed in a flow channel and adjusted through flow control system till system begins to oscillate. Once the system started oscillating readings were taken using the oscilloscope. The oscilloscope signal was adjusted to produce clear signal to measure peak to peak voltage on y-axis, and measure the time-period of the wave on the x-axis. Flow properties were adjusted to alter flow speed and submersion height, until system began to oscillate again. Once enough values were recorded for a given arm length the arm length was altered and tests were repeated.

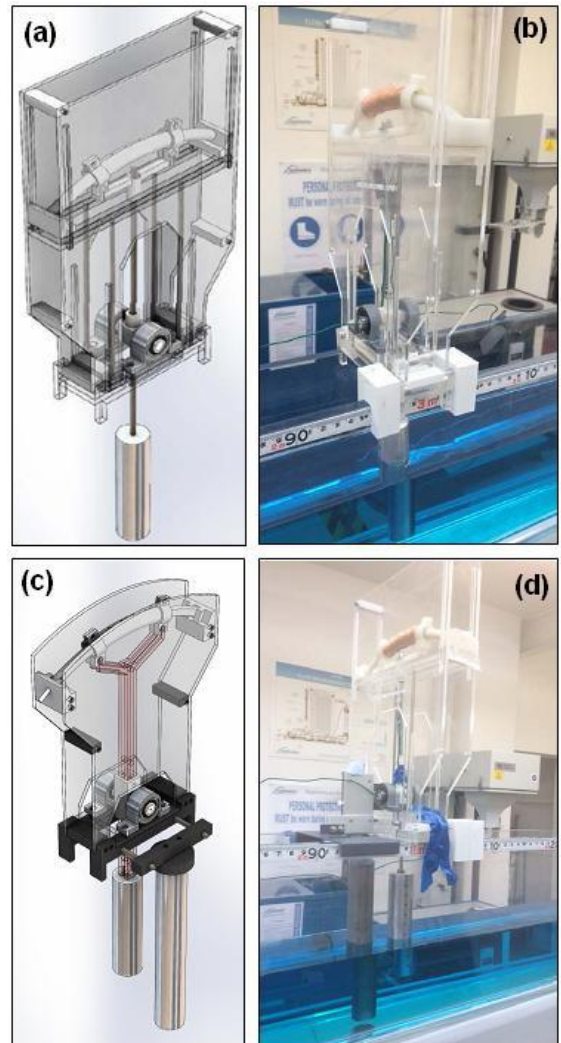

Fig. 6. Vortex-assisted generation system prototypes. (a) Design of the system without shedder, (b) Built prototype 1 without shedder, (c) Design of the system with vortex shedder, (d) Built prototype 2 with shedder.

The experiments showed that lock-in flow velocity and output voltage decreased with submergence, as depicted in Fig. 7. The arm length of 190 mm gives the highest voltages for the lower flow velocities. The arm length of 180 mm gives the highest voltages for the higher flow velocities. Higher lock-in flow velocities equates to lower output voltages. Two possible modes of vortex shedding modes have been observed with the arm length of 200 mm.

From Fig. 7 (a), a general trend can be seen that as submersion increases the output voltage decreases, this is consistent throughout all arm lengths. Lock-in flow velocity also decreases as submersion increases, independent of the modes, illustrated in Fig. 8. Looking further iit can be notices that the frequency is inversely proportional to the submersion. The arm length for the oscillator had very little effect on the natural frequency of the system. This was due to all lengths following the same trend in Fig. 7 (a) and (b). At shorter arm lengths, there is a higher voltage output at all velocities. From the results it was concluded that further work was essential in analyzing the different modes of synchronization, testing was carried out with a rigid shedder to identify possible optimization parameters.

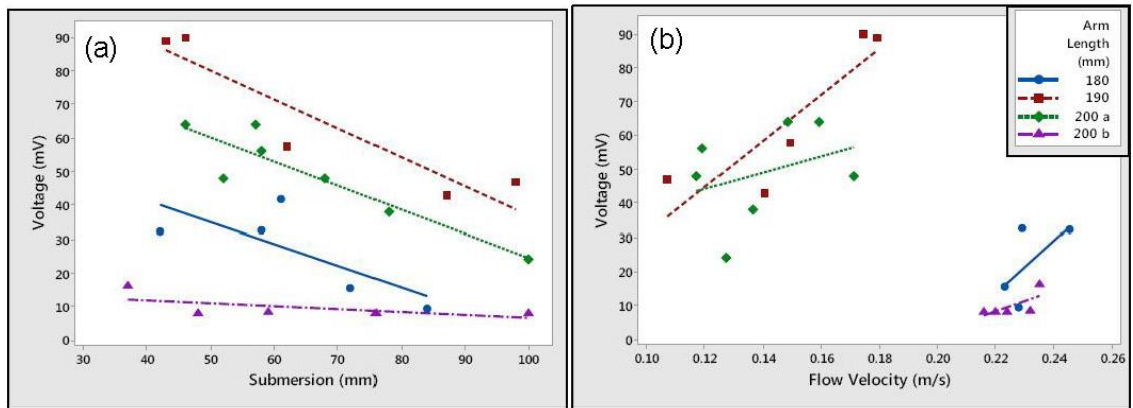

Fig. 7. Prototype 1 plots for (a) voltage vs. submersion and (b) voltage vs flow velocity for varied arm lengths

Table 1. Data collected from a single cylinder system testing

| Arm length (mm) | Submersion (mm) | Flow velocity (m/s) | Frequency (Hz) | Voltage (mV) |
|---|---|---|---|---|
| | 58 | 0.119 | 0.833 | 56 |
| | 52 | 0.171 | 1 | 48 |
| | 100 | 0.127 | 0.8 | 24 |
| | 78 | 0.136 | 0.83 | 38 |
| | 68 | 0.117 | 0.87 | 48 |
| 200 | 57 | 0.148 | 0.91 | 64 |
| | 46 | 0.159 | 1 | 64 |
| | 100 | 0.22 | 0.8475 | 8 |
| | 76 | 0.216 | 0.8197 | 8 |
| | 59 | 0.232 | 0.909 | 8.16 |
| | 48 | 0.224 | 0.862 | 7.92 |
| | 37 | 0.235 | 0.9259 | 16.2 |
| | 62 | 0.149 | 0.962 | 57.6 |
| | 87 | 0.14 | 0.93 | 42.8 |
| 190 | 98 | 0.107 | 0.87 | 47 |
| | 46 | 0.174 | 0.95 | 90 |
| | 43 | 0.179 | 0.95 | 89 |
| | 100 | 0.223 | 0.531 | 8.88 |
| | 61 | 0.145 | 0.935 | 41.6 |
| 180 | 84 | 0.228 | 0.892 | 9.28 |
| | 72 | 0.223 | 0.833 | 15.4 |
| | 58 | 0.229 | 0.909 | 32.4 |
| | 42 | 0.245 | 0.943 | 32 |

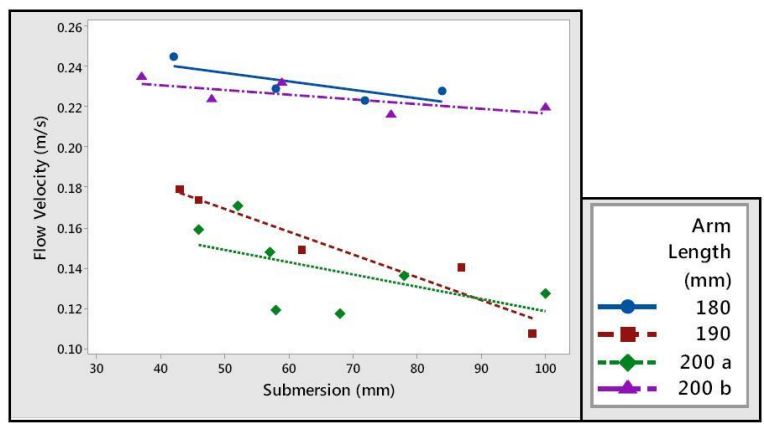

Fig. 8. Prototype 1 plots for flow velocity vs. submersion for varied arm lengths

Based on observations from testing for the second approach it was noted that using a bluff body at a given distance from the oscillating bluff body, larger oscillation amplitudes could be produced. More importantly the oscillations occurred consistently at various flow rates. The objective for the second testing was to examine the effect of using a bluff body as a vortex shedder to enhance energy generation allowing for comparison to direct vortex induced vibration system explained above. The lower arm radius length was adjusted and shedder body was set to a distance from the oscillating body. The flow velocity and submersion depth of oscillating cylinder were set. Peak to peak voltage and frequency of generated oscillations were recorded. Shedder distance was adjusted and tests were repeated. The submergence was adjusted but the flow velocity was kept at same rate and tests were repeated. As the second prototype incorporates two cylinders, it is important to define which cylinder is which. In this report the vortex shedder or splitter is the rigid body in front of the device that sheds the vortices. The bluff body is the cylinder which is free to move in the direction normal to the flow and oscillates due to the shed vortices.

Raw data was recorded for multiple set-ups, where the submersion and, shedder to bluff body spacing were varied to identify their effects on the system as shown in Fig. 8 and Table 2. As flow velocity increases the following increases: output voltage, frequency, power output, mechanical extraction power, system efficiency. As flow velocity increases, extraction efficiency decreases. As shedder distance increases the extraction efficiency increases, whilst the system efficiency decreases. Lower flow velocities have higher efficiencies with higher submersions. Extraction efficiency does not seem to be effected by submergence. A transition point occurs around 0.21 m/s where at lower velocities a lower submersion is favored to yield higher output voltage and frequency. The spacing of the shedder to the bluff body is inversely proportional to the voltage and frequency output

Fig. 9 (a) shows that output voltage increases as submersion level increases, for the experimental range of flow velocities. Lift forces acting on the oscillating body is related to the integral of the pressure distribution across the stream wise length. By increasing the submersion level, and thereby the fluid contact surface area, the lift forces effect is larger. Furthermore, a general observation is that as the flow velocity increases, the frequency increases, however the submersion is independent of the frequency. In addition the coefficient of lift is the highest at the higher flow velocity, 100 mm submersion and 25 mm shedder spacing when the frequency remains approximately constant as shown in Figs. A.1, A.2 and A.3 in Appendix A. A higher submersion has a smaller measured range of active flow velocities as seen in Fig. 9 (a). Submersion has a very small/no effect on the extraction efficiency. For the same flow velocity a higher submersion has greater output power and system efficiency. This is explained by the trend where a larger submersion has a higher input power to the system at constant velocity; this, coupled with the similar extraction efficiency, leads to an increased power output.

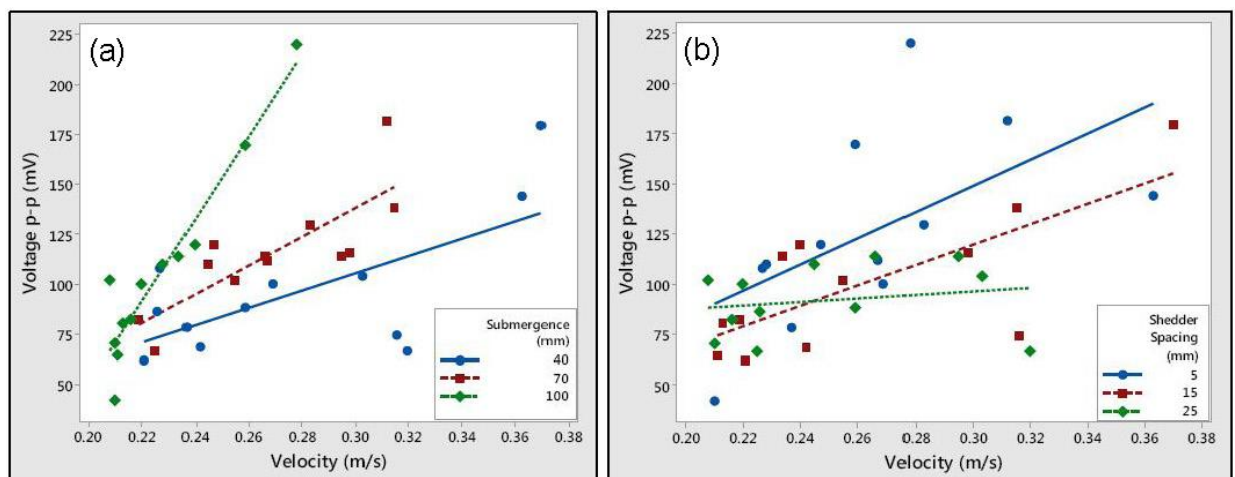

Fig. 9. Prototype 2 plots for (a) voltage vs. velocity for varied submergence and (b) varied shedder spacing

From the analyses it has been found that the lift forces acting on the oscillating cylinder are influenced by the shedder spacing. The vortex dissipation and development process could be used to explain the maximum lift generated and it can also be used to explain the extracted efficiency of the system. Fig. 9 (b) displays that as the spacing of the shedder to the oscillator is increased, the output voltage lowers. With accordance to this observation, this is possibly due to dissipation of energy from the vortices as they move further downstream from the shedder, due to natural decay. This hypothesis suits the gathered data for all flow velocities over 0.25 m/s; where there is more energy contained in the vortices at 5 mm from the shedder than at 25 mm.

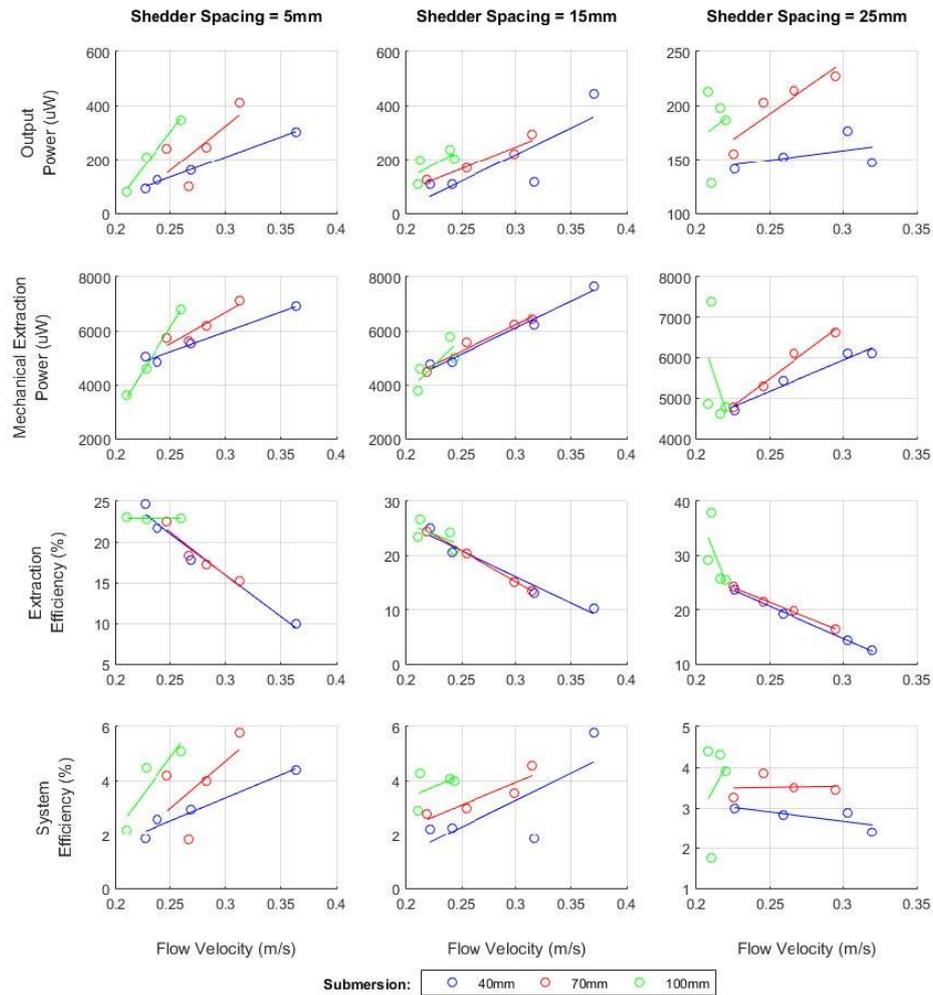

Fig.10. Prototype 2 plots for power and efficiency

From Fig.10 and Figs. A.4-A.9 in Appendix A, it is observed that shedder spacing has no significant observed effect on the output power, mechanical extraction power, extraction efficiency, and system efficiency. There is a contradictory case to this observation for 25 mm shedder spacing at 100 mm submersion. This could be a result of the non-uniform flow velocity profile in the channel, where the lower (slower) plane of flow will have a lower shedding frequency, and therefore produce vortices at a different rate to higher planes in the channel. If the shedder spacing is such that the difference in plane vortex formation causes interactions, the overall transverse lift forces acting on the oscillator will be affected. This is a complex case and requires significant extra work to understand fully.

For all cases in terms of submersion and shedder distance, and increase in velocity lead to a larger voltage output. This is possibly due to the higher energy in the flow. From Figs. A.10-A.15 in Appendix A, a distinct transition point occurs around 0.21 m/s where lower velocities at lower submersions have a positive effect on the output voltage and frequency. This could be due to the slower flow rate, decreasing the shedding frequency, and with a lower surface area resulting in increased synchronization of the vortices in the vertical axis of the streamline planes. Fig. 10 shows for all cases that the power output, mechanical extraction power, and system efficiency increases as flow velocity increases. Note there is one contradiction to this at shedder spacing of 25 mm where the system efficiency appears to decrease. Fig.10 also shows that as the flow velocity increases the extraction efficiency decreases. This is due to the large increase in input fluid energy, proportional to the cube of the velocity, relative to the linear increase in mechanical extraction power, resulting in a decrease in extraction efficiency by proportional difference.

Table 2. Data collected from testing a system with vortex shedder and bluff body

| Lower arm radius (mm) | Submergence (mm) | Shedding distance (mm) | Velocity$_1$ (m/s) | Voltage$_1$ (mV) | Frequency$_1$ (Hz) | Velocity$_2$ (m/s) | Voltage$_2$ (mV) | Frequency$_2$ (Hz) | Velocity$_3$ (m/s) | Voltage$_3$ (mV) | Frequency$_3$ (Hz) |
|---|---|---|---|---|---|---|---|---|---|---|---|
| 180 | 100 | 5 | 0.278 | 220 | 1.761 | 0.259 | 170 | 1.337 | 0.21 | 42 | 0.7143 |
| | | 15 | 0.211 | 64 | 0.7396 | 0.24 | 120 | 1.136 | 0.234 | 114 | 0.9843 |
| | | 25 | 0.21 | 70 | 1.453 | 0.22 | 100 | 0.9399 | 0.208 | 102 | 0.9542 |
| | 70 | 5 | 0.247 | 120 | 1.126 | 0.283 | 130 | 1.214 | 0.312 | 182 | 1.404 |
| | | 15 | 0.298 | 116 | 1.225 | 0.255 | 102 | 1.096 | 0.219 | 82 | 0.8803 |
| | | 25 | 0.245 | 110 | 1.042 | 0.266 | 114 | 1.202 | 0.295 | 114 | 1.302 |
| | 40 | 5 | 0.237 | 78 | 0.9542 | 0.363 | 144 | 1.359 | 0.269 | 100 | 1.087 |
| | | 15 | 0.37 | 180 | 1.506 | 0.242 | 68 | 0.9542 | 0.316 | 74 | 1.225 |
| | | 25 | 0.226 | 85 | 0.9259 | 0.259 | 88 | 1.068 | 0.303 | 104 | 1.202 |

## 6. Computational analysis

For the computational analysis of the system a variety of CFD methods were applied. Vortex formation is known to be a transient phenomenon; therefore, all of the solutions were defined to be in a transient domain. The methods presented below increase in complexity, accuracy, and necessary computational resource as the list continues. It was necessary to identify all the means of analysis so that the most appropriate could be used for the final analysis of the system. The computational analysis was performed in parallel to the empirical, with an extended range of flow velocities, to investigate the flow velocities before and after the apparent transitional point. The oscillator submergence was set to 100 mm, 70 mm, 40 mm, the shedder spacing was set to 25 mm, 15 mm, 5 mm, and the flow velocity was set to 0.28 m/s, 0.21 m/s, 0.14 m/s

### 2D static domain

The simplest analysis possible to understand vortex formation is a two dimensional static analysis. The domain was created in ANSYS Geometry Modeller and loaded into Fluent, where the boundary conditions were defined and the solution was calculated. Fig. 11 shows a typical solution, where a clear von Karman vortex street is seen, and also the theory is confirmed by the difference in the pressure values around the bluff body stream wise length.

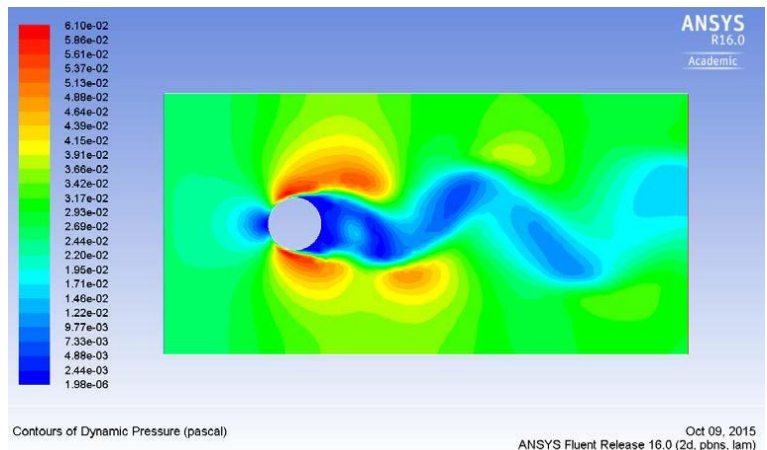

Fig. 11. Dynamic pressure distribution for the entire solution domain

### 2D dynamic domain

A dynamic analysis of the two dimensional domain can be created using the same method as the static analysis and incorporating a user defined function (UDF) that handles the movement of the body and dynamic meshing. The UDF tells the domain that the bluff body has freedom to move in the desired axis, and has a mass (therefore inertia). When the force on the body is such that it will move, the domain will re-mesh according to the set parameters to ensure a smooth transition with minimum loss of accuracy of the element discretisation. Fig. 12 shows a typical example of this at two time points from the results generated in Fluent. This is a good way to learn about the nature of practical VIV motion. The normalized values for coefficient of lift and displacement are shown in Fig. 13.

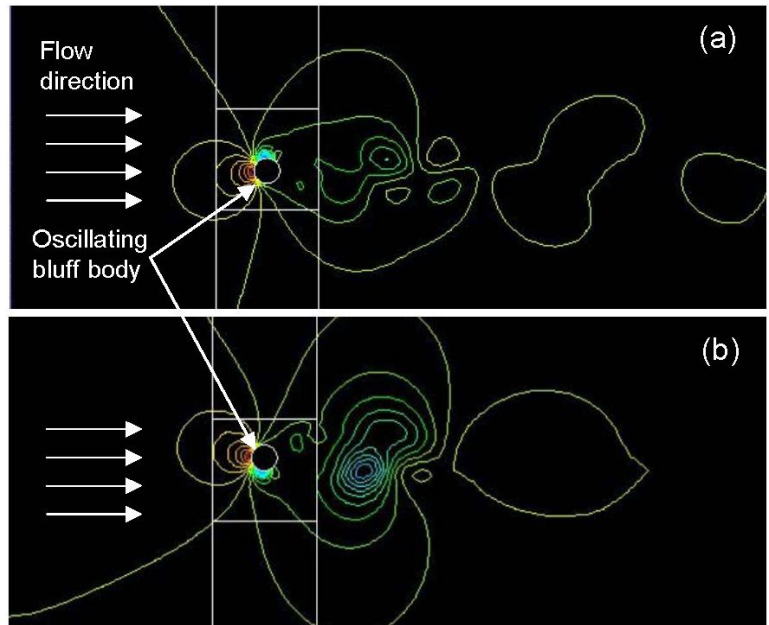

Fig. 12. Pressure distribution in solution domain at different times. (a) 4.5 s, (b) 5.0 s

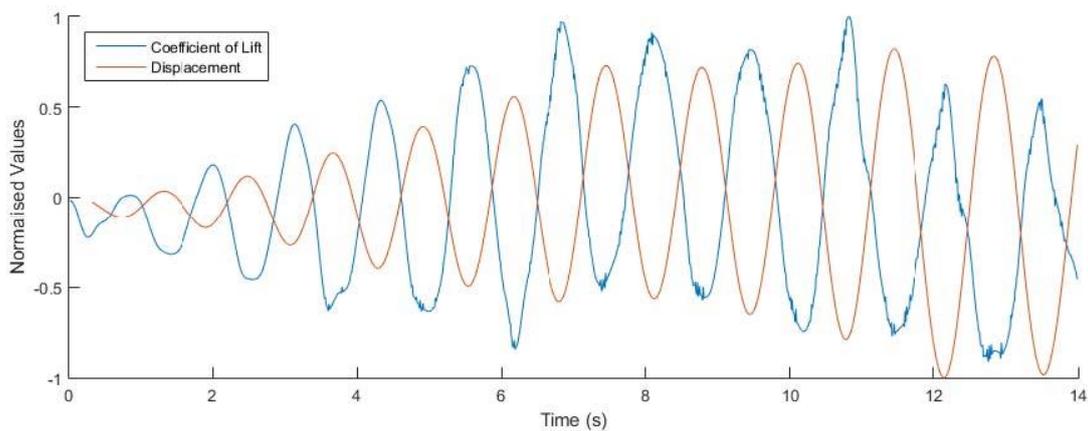

Fig. 13. Normalized values for coefficient of lift and displacement of 2D dynamic domain

### *3D static domain*

For this method three dimensional models were created using SolidWorks and imported into ANSYS Workbench for mesh generation. The mesh files were then exported into Fluent where the boundary conditions were set similarly to the previous methods. A large number of models can be quickly processed this way, as the solution considers a simple geometry rigid domain. Post-processing was either performed using Fluent or the dedicated CFD-Post application provided by ANSYS as shown in Figs 14 and 15. This provides a multitude of information that is helpful in the analysis of the system. It is seen in the Fig. 14 (b) that the vortex formation is more distinct when the vortex shedder is used and it causes the symmetry breaking that eventually leads to the von Karman vortex street pattern. The asymmetry in the wake continues to grow from there until shedding occurs. Otherwise the wake would remain symmetrical and vorticity would be diffused from the wake, as opposed to being shed, as shown in Fig 14 (a). For the cylinder undergoing forced oscillations, the characteristic formation time marks the switch in the phase of the wake, relative to the motion of the cylinder.

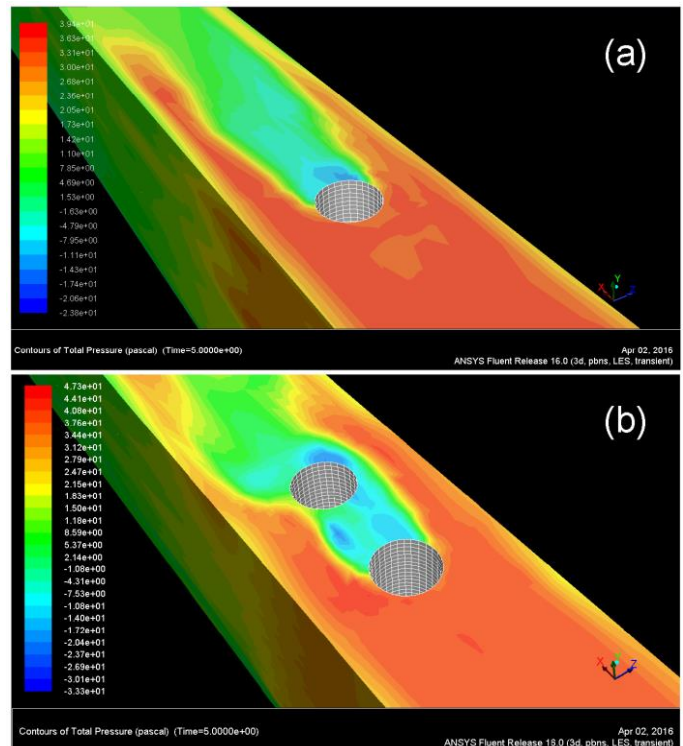

Fig. 14. Fluid pressure contours (a) without shedder and (b) with shedder

Fig. 15 and Figs. B.1 and B.2 in Appendix B show the pressure distribution on a bluff body and absolute velocity helicity for planes of fluid. Helicity is a scalar quantity defined as an inner dot product of velocity and vorticity vectors. As such, the helicity can be used as a useful indicator of how velocity vector field is oriented with respect to vorticity vector-field for a given flow field.

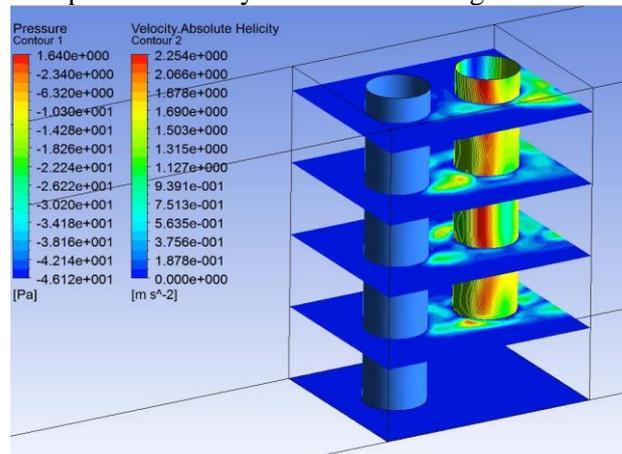

Fig. 15. Pressure distribution on a bluff body and absolute velocity helicity for planes of fluid

### *3D dynamic domain – Fluid Structure Interaction*

Fluid-structure interaction (FSI) utilizes transient mechanical and flow analyses joined by a system coupling in the ANSYS Workbench. Similarly to the three dimensional static method, the model was created in SolidWorks, Fig. 16 (a), and imported into ANSYS. However in this method the solids and fluids were meshed and analyzed independently, as can be seen in Fig. 16 (b) and (c)-(e). The transient structural setup defines the dimensional freedom of the body and the forces in the form of a fluid-solid interface. The Fluent setup is the same as the previous method, with a dynamic mesh set to the fluid-solid interface. The solution is iterative by time step with information sharing between the two domains of structural and fluid. Initially the mechanical solver computes the solid domain solution, then Fluent computes the fluid domain, and the loop carries over. The mechanical interprets the influence of the fluid on the body, and applies equivalent forces which result in body motion. Fluent interprets the motion and re-meshes the domain appropriately to ensure continuity. This process occurs every iteration of every time step. Ultimately this is a very slow and labor intensive computation, even at very low mesh refinements and simple solution solver methods. The outputs for this method are much more realistic than any other analyzed in this report, as it shows actual movement of the bluff body due to the motion of the fluid.

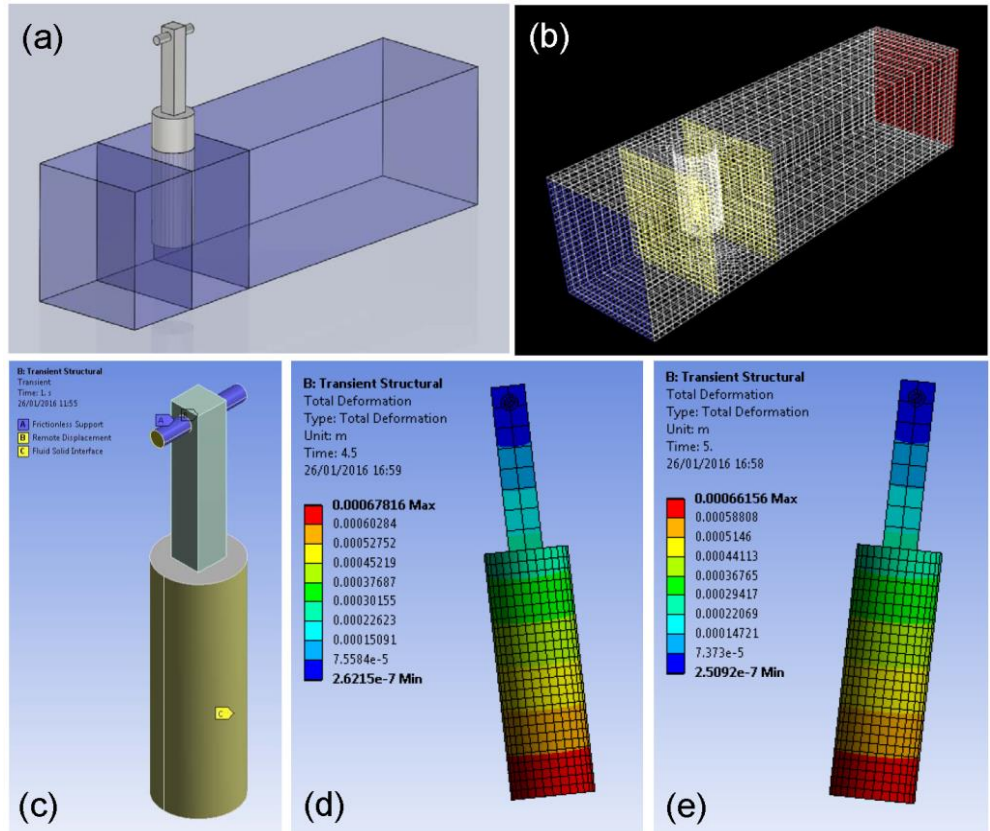

Fig. 16. 3D dynamic fluid-structure interaction analysis. (a) 3D CAD domain, (b) mesh with boundary regions, (c) transient structural constraints on pivoted cylinder, (d) and (e) total deformation of the pivoted cylinder at different times.

Fig. 17 shows the fluid force distribution around the pivoted cylinder obtained at different times of 4.56 s and 5.0 s. It is seen that the force in front is acting on the cylinder and its magnitude varies on the left and right hand sides of the cylinder at different time increments. The figure below shows the acceleration vectors pointed to the same direction of the acting force. This analysis allowed determining the direction and magnitude of force acting on the bluff body to evaluate the system's performance. The simulation results show that the numerical method is able to reproduce the vortex-induced vibration phenomena.

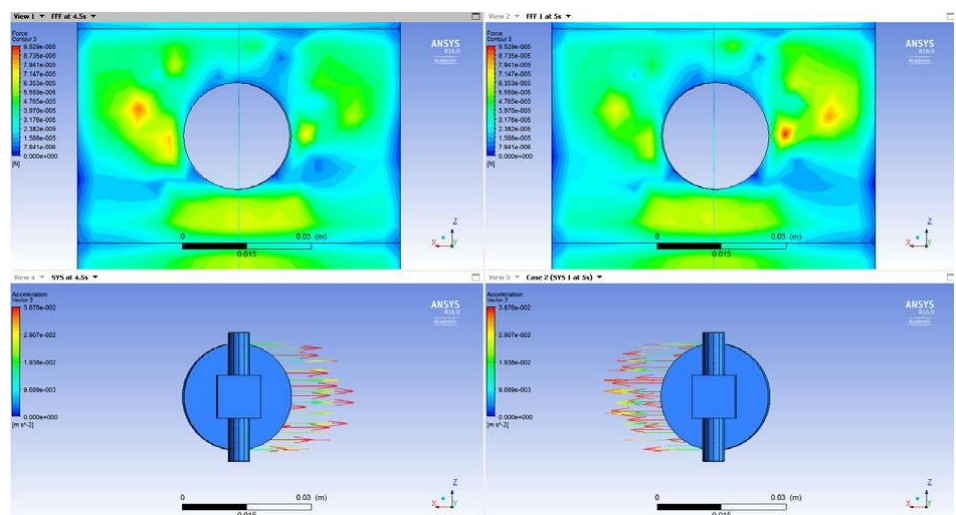

Fig. 17. Fluid force contours and pivoted cylinder acceleration vectors at different times

Fig. 18 shows the results of normalized displacement, velocity and acceleration over a time period. The maxima for the displacement and acceleration have the same trend and the one for velocity shows the opposite trend. These are in agreement with the results shown in Fig. 13, where the coefficient of lift and displacement showed an opposite trend.

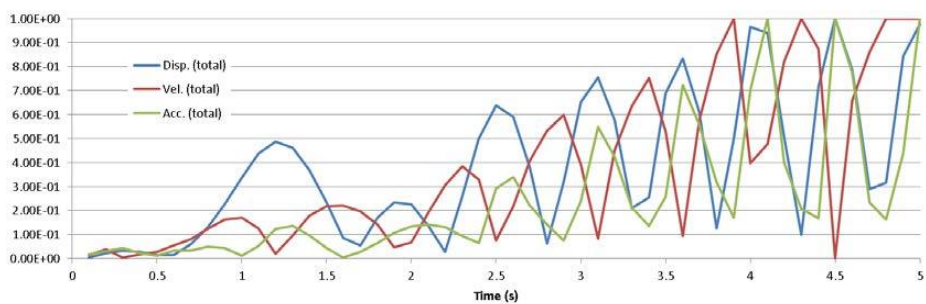

Fig. 18. Normalized system displacement, velocity, and acceleration vs. time

Vortices made by the shedder body are irrational, and the particle velocity inside the vortex decreases from the axis line. From Bernoulli's principle, it is possible to show that the dynamic pressure is proportional to the square of the radial distance from the vortex axis line. The pressure and movement of the vortex forces the fluid to move the oscillator cylinder. The higher the total dynamic pressure in the vortices, the higher the cylinder lift coefficient. A moving vortex inside a medium has angular and linear momentum which transfers both energy and mass. In an ideal fluid this energy can never be dissipated, but for a real fluid there is a dissipation of energy due to fluid viscosity, and the regular pattern of the vortex von Karman street gradually fades. Extensive research concerning this has been carried out by Akhmetov [41], where he theoretically discussed the energy dissipation in a medium under the influence of viscosity.

Similar research was conducted by Sennitskii [42], who demonstrated that when two cylinders are placed in a parallel formation, the total drag of the moving cylinder is always lower than that of stationary cylinder. He also investigated how the development process affects the properties of the vortices. It is therefore possible to explain the CFD and experimental data by considering the energy development and dissipation during the motion of a vortex. The vortices are generated on both sides of the cylinder with a specific shedder frequency and the second cylinder forces them to join. It is the product of the two formed vortices that move around the second cylinder to generate a lift force as shown in Fig. 19. By considering the size of the vortices and the stored energy within a vortex with its dissipation rate for a 2D system, it is possible to justify the reasons for a decrease of maximum output power when the shedder spacing was increasing. However the majority of the investigations were carried out in 3D domain and the submersion of the cylinder is a major factor that drastically influences the results. As depicted in the Fig. 20, the submersion of the cylinder changes the vertical profile of the generated vortices. The colour contour on the cylinder is showing the pressure which is a function of the lift forces and the vortices are shown by the absolute helicity in the fluid velocity. This interesting phenomenon is very complicated and requires further investigation to fully understand.

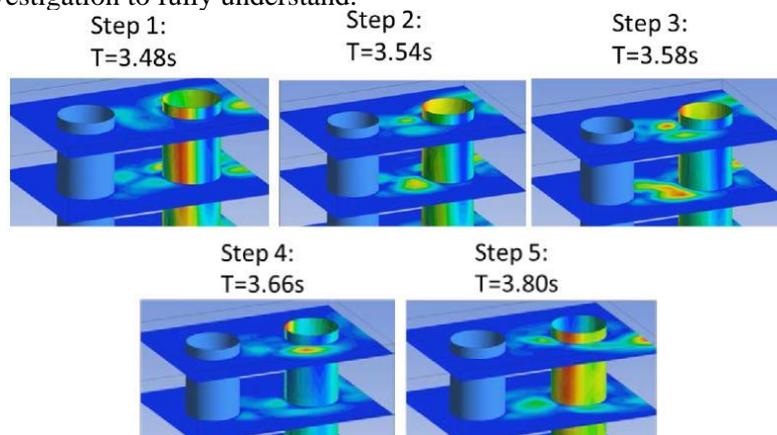

Fig. 19. The convergence of two vortices producing a local transverse lift force acting on the oscillating body over time

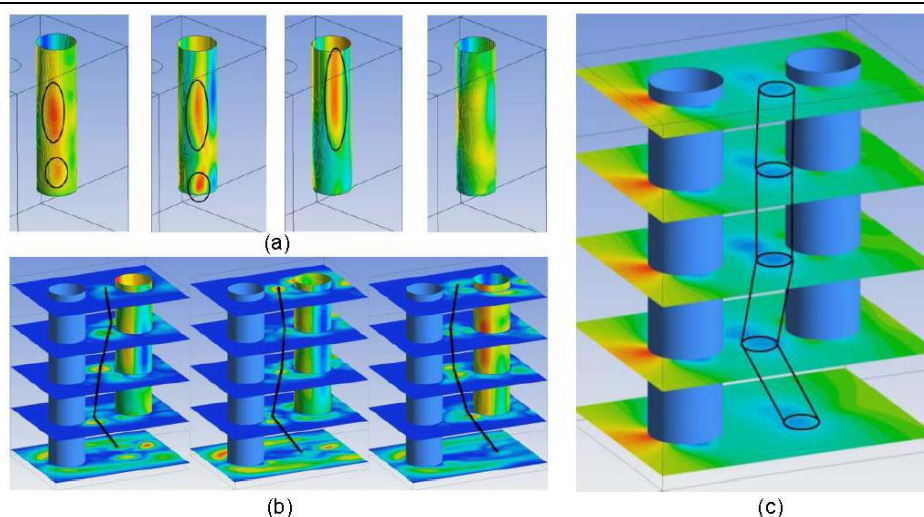

Fig. 20. Pressure distribution acting on (a) the second cylinder, (b) curved vortex profile by fluid velocity helicity, and (c) curved vortex profile by fluid pressure.

As seen in Fig. 20 the curvature of the vortex is generating local lift forces which vary along the length of the cylinder. This variation is seemingly chaotic, Fig 20 (b). A possible explanation for the curvature is because of the effect of the boundary layer from the flow channel surfaces, as well as the separation of the flow from the underside of the second cylinder.

## 7. Potential Application Examples

The consumption and pollution of water by agriculture is becoming a serious global concern. The V.A.G. system produced in this project can operate in very slow moving waters, making it a viable option to use in the water channels in farms aiding the efficiency of a farmer's water usage. Multiple V.A.G. units can be placed in series formation where the generated electrical energy can be harnessed in storage units. For example, rice is one of the major crops feeding the world population and farmers in countries with low water supply can reduce their farm's water-footprint by controlling the water flow to crops.

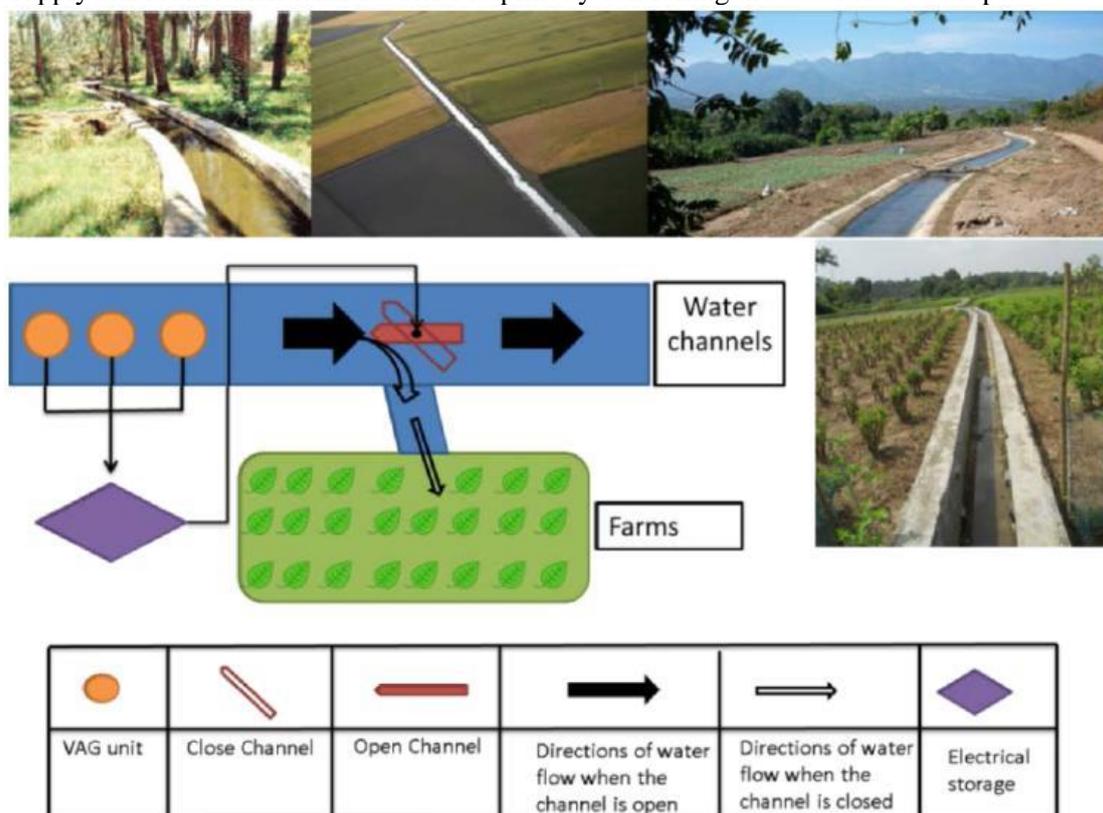

Fig. 21. Potential application of V.A.G. system for rice farm irrigation

Some specific crops such as Sorghum, which is used for food, both animals and humans, and for ethanol production, only need a limited water supply and the proposed system can be used to manage the water flow to these farms [43]. Fig. 21 shows the application of V.A.G. in such farms and in this case it is used to open and close the water channel gates to control the water flow.

The V.A.G. system can also be used directly to generate electricity from slow rivers or canals and the system can be scaled up as a group of devices localized in a required space or area, as shown in Fig. 22. The system can be very easy installed and maintained and does not make any disruption to the flow. Another advantage of the system is that the generator and any electrical parts and joints are located outside the water and not affected by short circuit and corrosion.

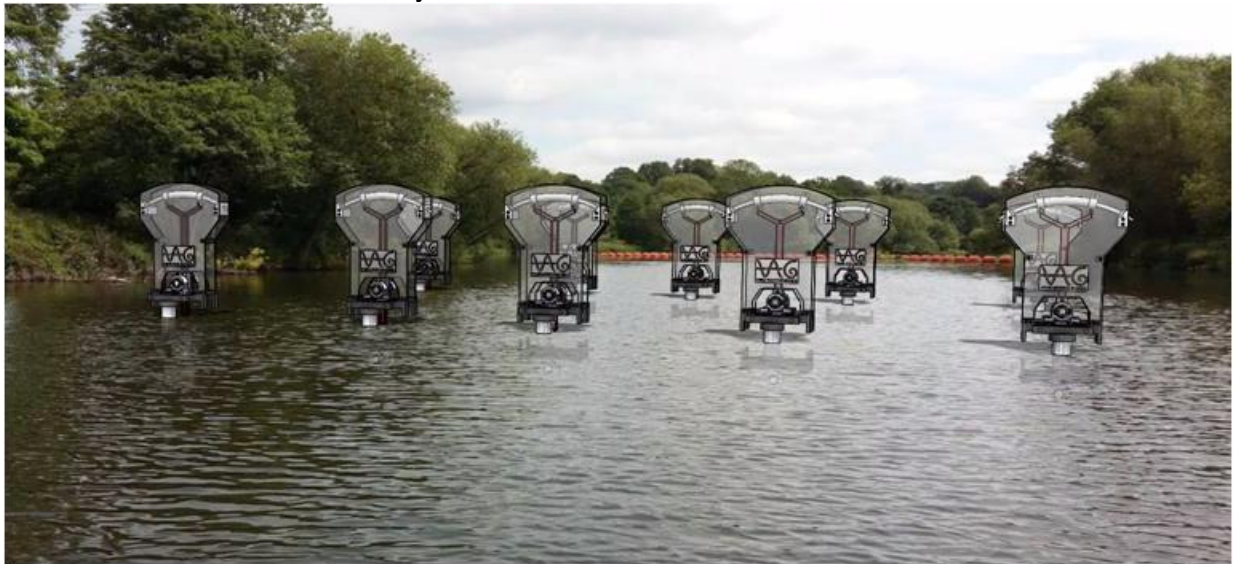
Fig. 22. Potential application of V.A.G system in a slow-water river

## 8. Conclusion

Two systems were produced that utilize VIV to generate energy. The methods in which they cause oscillation of the pivoted cylinder differ slightly. The first method functions by direct induced vibration of the bluff body by using its geometry to induce the vortices causing the vibration. The second system uses a shedder (rigid bluff body) that produces vortices to generate oscillations of the pivoted bluff body behind the shedder.

The flow velocity was taken at a set point in the flow channel, downstream of the system to ensure it did not interfere with the flow before the bluff body. It was set over a meter downstream of the system, which allowed for a fully developed flow profile to develop. However due to the distance the drop in the average velocity is not accounted for, which is expected due to the viscous effects of the channel on the fluid boundaries. This is acceptable for system comparison, as it was in the same position for all experiments. Multiple readings were taken of the flow speed and an average of the flow speed was used to increase the accuracy and reliability of the results. It was important to keep the flow conditions constant for every experiment and to ensure the flow was fully developed, the system was placed 3 m away from the source of flow. This was an essential factor because maximum lift is generated when the flow is fully developed.

Accuracy and reliability of recorded results was also an important consideration. An oscilloscope was used for the experimentation which allowed data to be directly recorded in the form of .csv files. This was then able to be processed in Matlab. Efforts were put into achieving a clear signal during testing. Given that the output of the oscillations required analysis and was not just concerned with energy generation efforts were put into minimizing the effects of interference. From initial observations the systems alignment was an issue, where the coil ring and stator's small clearance prevented optimum, fluid oscillations. Through model development and appropriate lubrication, this affliction for friction has been minimized to create a more robust and effective system.

### *Prototype 1*

From the tests carried out on the Prototype 1, the natural frequency (~0.9Hz) of the oscillating system was able to be identified. For any submersion and speed the system oscillates at a definitive frequency.

This relates to the theory where lock-in can only occur when the shedding frequency is equivalent to ±20% of the natural frequency of the system. This explains why Prototype 1 was only able to oscillate at definite flow velocities and submersion depths, which makes it unreliable in application as the requirements for VIV is very dependent on the flow conditions.

Two possible modes of the lock-in were observed. This was identified by the two different lock-in flow velocity ranges for the same arm length case of 200 mm. The cause is likely to be the difference in shedding frequencies while the oscillation frequency remains constant at these flow velocities. From these two modes, at lower velocities a higher output voltage is generated, whilst at higher flow velocities, a more reliable output is achieved at steady flow rates with a varied submersion. This is due to the shedding frequencies at lower flow rates being closer to the natural frequency of the system. Whilst the system tends to oscillate about its natural frequency, a higher voltage is produced the closer the shedding frequency is to the system's natural frequency. Prototype 1 relies heavily on the conditions of lock-in being satisfied for energy generation which has proven to be unreliable; therefore power measurements were not considered for this system.

## Prototype 2

System two utilizes a shedder in-front of the oscillating bluff body. This was developed after it was noticed that by having the oscillator in the wake of the shedder bluff body, oscillations were able to be produced more consistently and reliably. At flow conditions where the first system would not oscillate; oscillations could be made possible using the shedder. This was significant for the implications that this had on possible applications, meaning that the system could now be placed in inconsistent conditions.

Important observations were made in means of set-ups of the systems for given applications. It was found that the spacing between the two bluff bodies was inversely proportional to the voltage; hence for maximum generation an application would favour a smaller spacing. In addition it was seen that voltage output was proportional to the submergence of the system, hence applications would favour the system to be submersed as much as possible. A transition point in velocity was also found at a flow velocity of $\approx 0.21 \text{m/s}$. Beyond this point it could be seen that a higher submergence level would give you a higher output voltage, however below this value the system favours a lower submergence. Further evidence of the effects contributing towards performance defining parameter was investigated by a CFD analysis.

## System Comparison

As can be seen from the analyses of the two prototype systems, the second prototype offers the best functionality, especially in how utilizing a shedder allows reliable and consistent energy generation in varied flow velocities and submersion levels. This is especially important in the implications that this has on possible applications, wherein Prototype 1 is limited by the requirements of lock-in to be satisfied to be able to oscillate, which hinders the use in applications where flow rate will vary.

The comparison shows that Prototype 2 generates higher voltages and frequencies at all submergence levels when compared to Prototype 1. This is further supported by the CFD analysis where it is seen that incorporating a shedder increases the coefficient of lift experienced on the oscillator. Further to this it was found that the addition of the shedder causes the wake to be amplified.

The potential for application of the developed system is promising, as there are many real world environments that similar to the testing conditions. The system could be used to store electrical energy, or power a minute electrical device. Alternatively, it could be used to measure flow velocity, if the other system parameters are known. If future work proves this as an accurate method, it would be a highly cost effective device compared to other less environmental friendly products, featured on the market today.

There is a large scope for further work, for example, the one concerning the scalability of the V.A.G. system, it may be applicable for use in harsher operating conditions such as large rivers. The phenomena of shedder induced VIV could be further investigated by computational methods, so that a means of optimization may be established. This may result in an overall design methodology for the system.

The final deliverable of this project is a novel system that operates at good extraction efficiencies (up to 37%). Prototype 2, alternatively known as V.A.G. operates in slow currents below 1 m/s. This registers a maximum power output of 0.45 mW at 0.36 m/s and with further work on the electrical magnetic generator to amplifying this, if scaled up, could be potentially used as an efficient renewable energy harvesting technique.

**APPENDIX A**

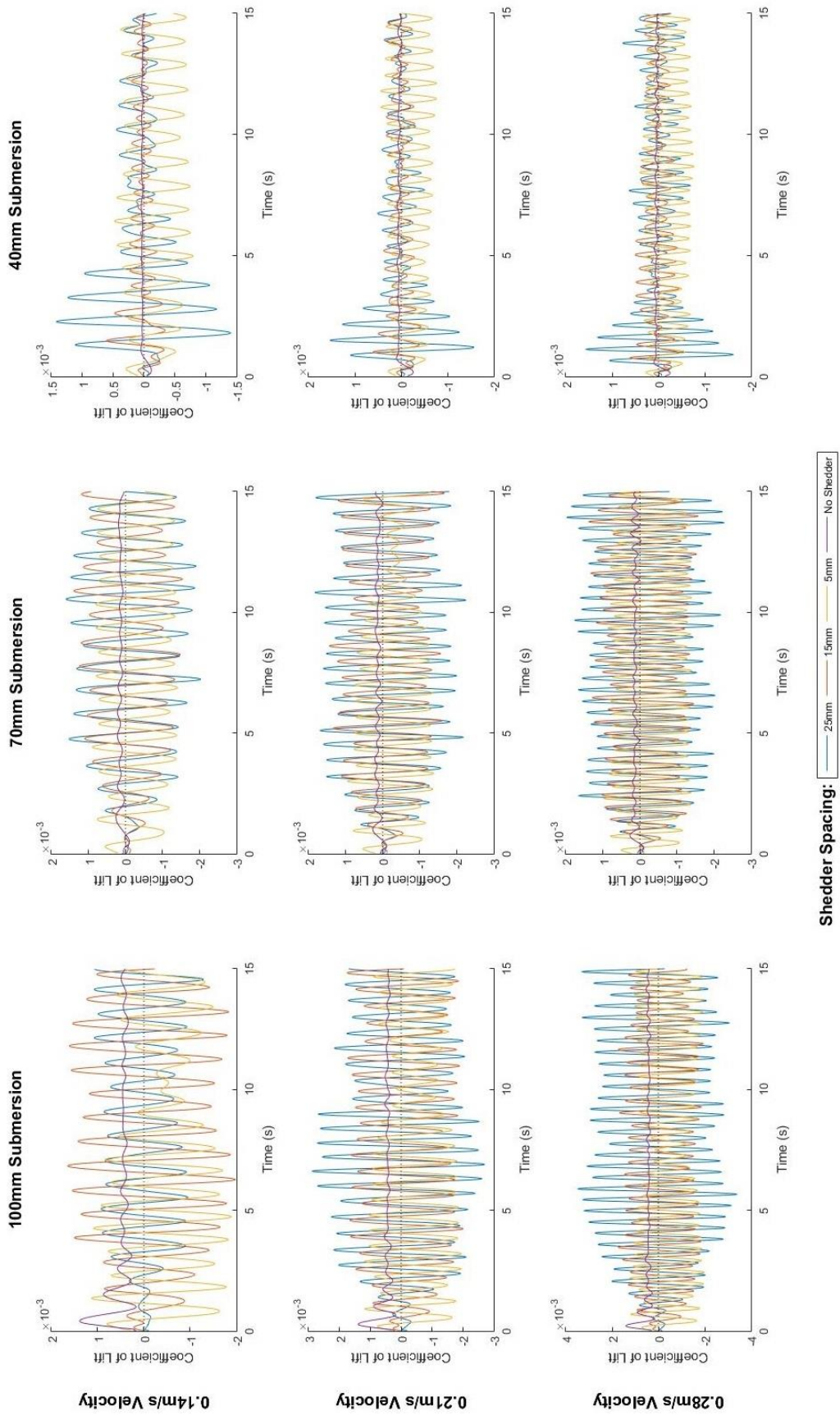

Fig. A.1. Submersion and velocity matrix of plots for Coefficient of Lift vs. Time for varied shedder spacing

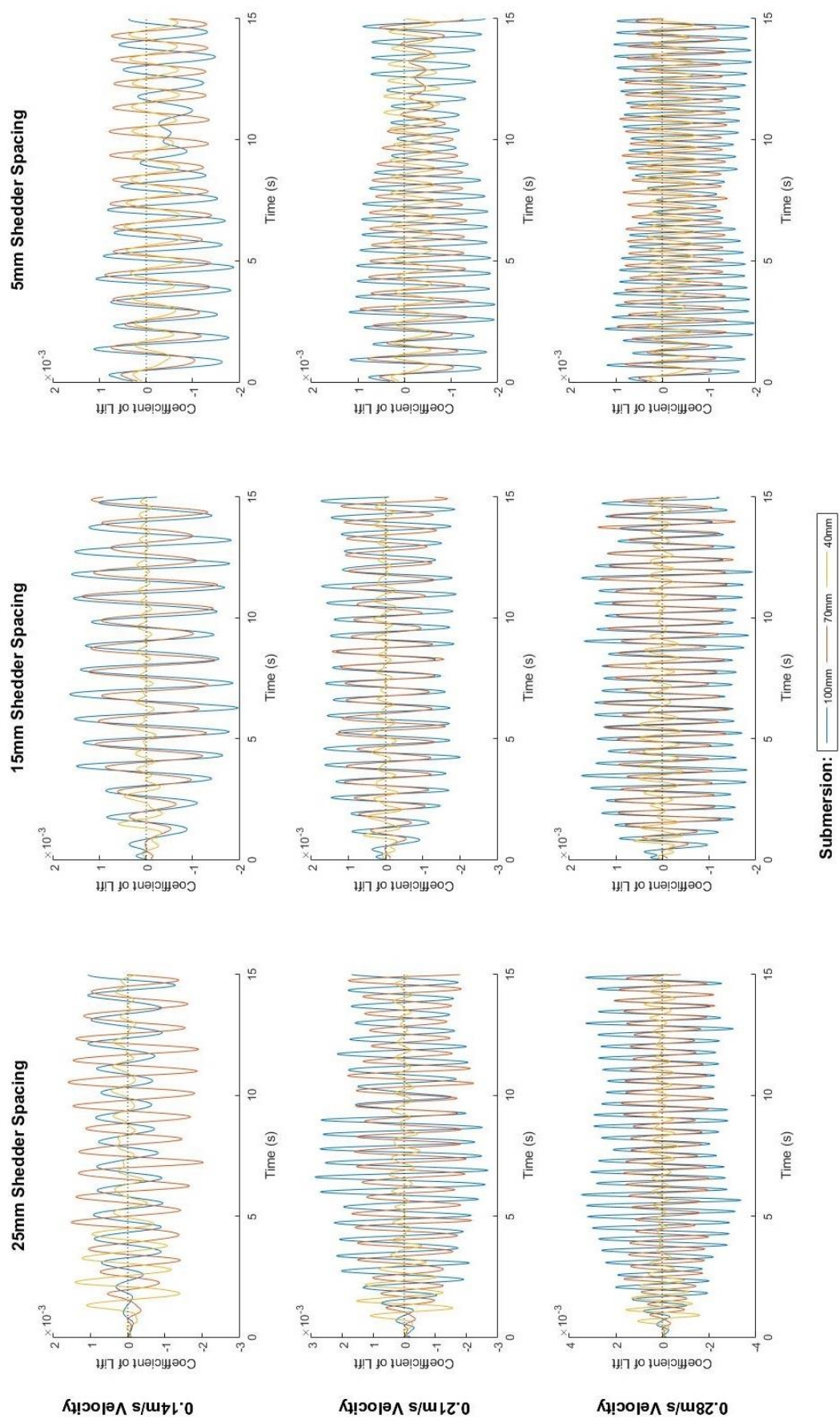

Fig. A. 2. Shedder spacing and velocity matrix of plots for Coefficient of Lift vs. Time for varied submersion

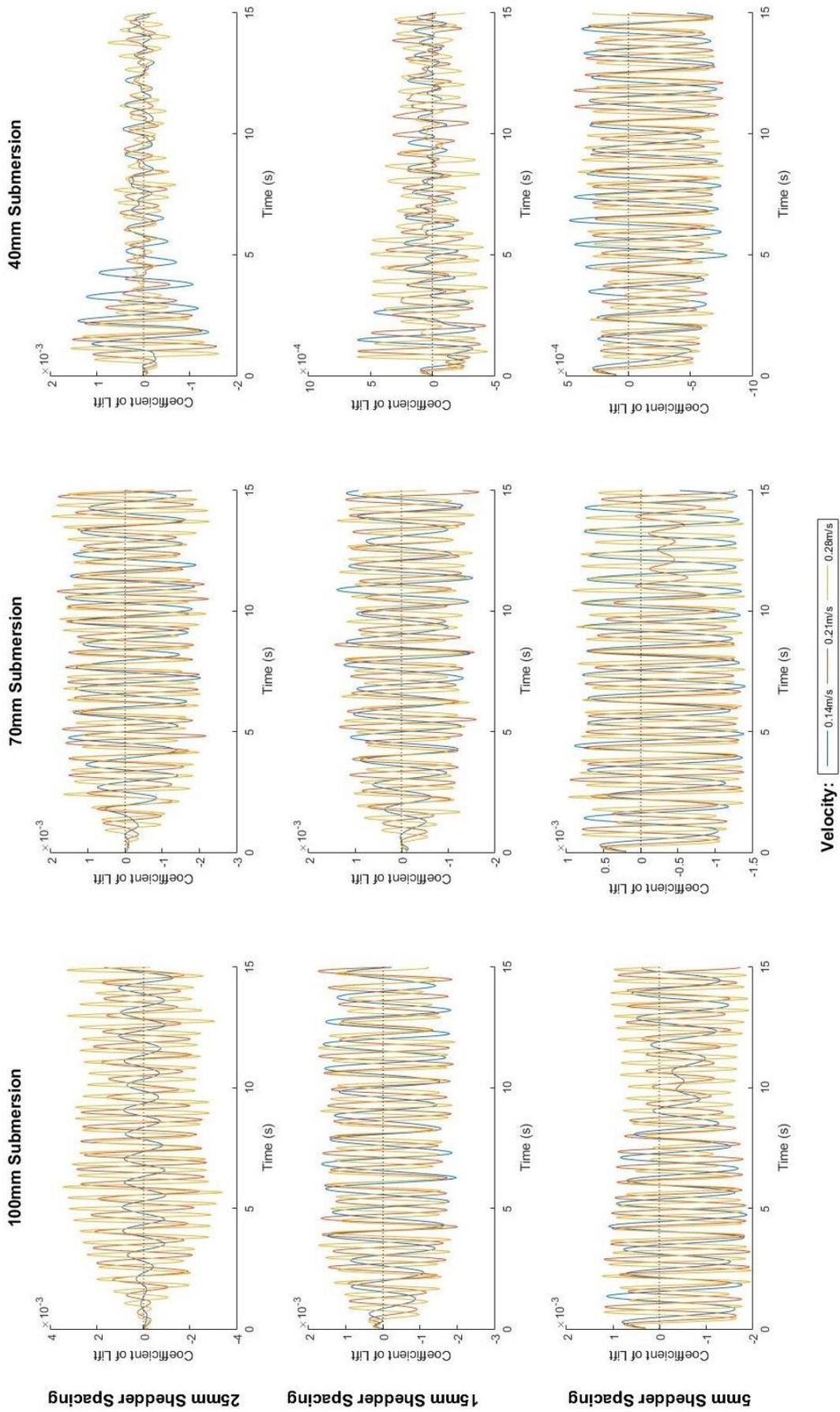

Fig. A.3. Submersion and shedder spacing matrix of plots for Coefficient of Lift vs. Time for varied velocity

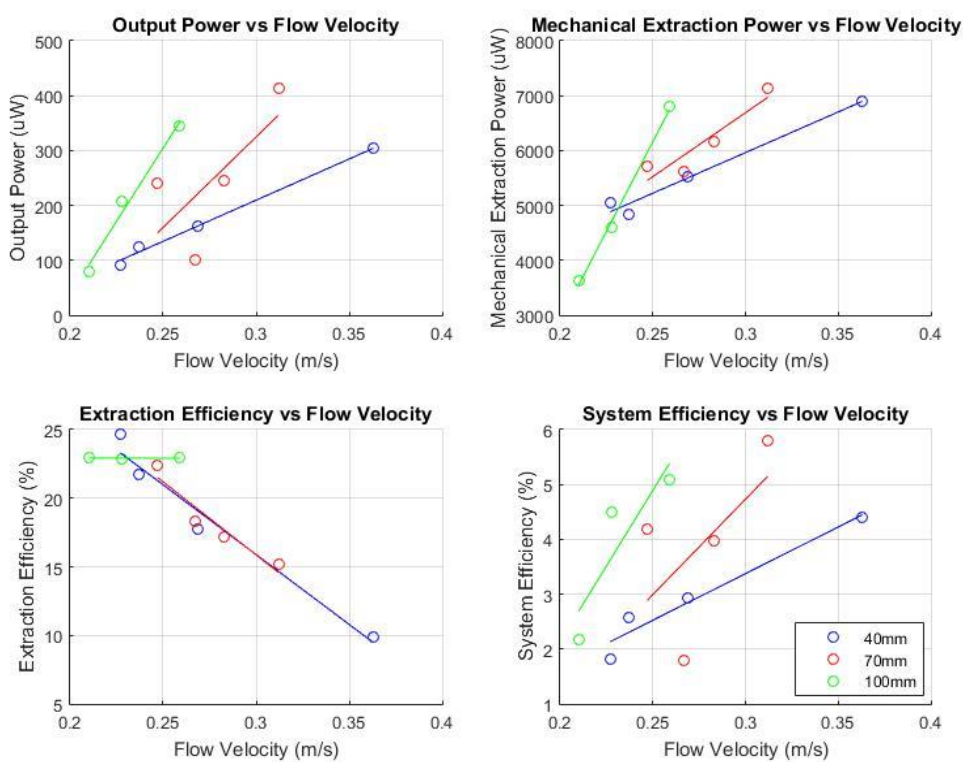

Fig. A.4. Power and efficiency for shedder distance 5 mm

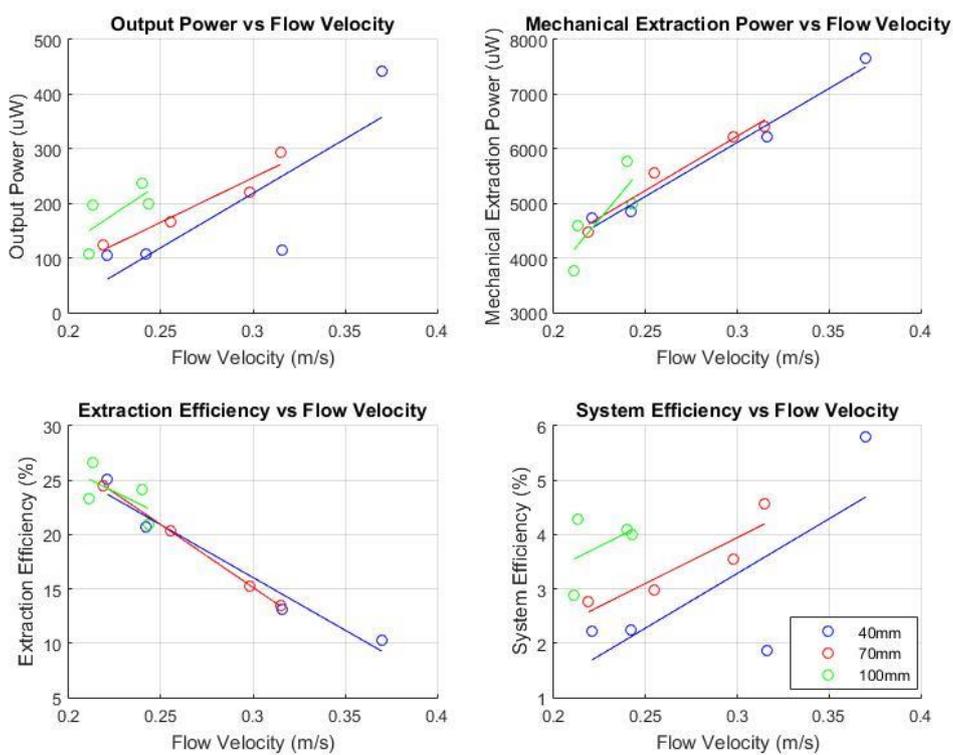

Fig. A.5. Power and efficiency for shedder distance 15 mm

Splitter Spacing = 25mm

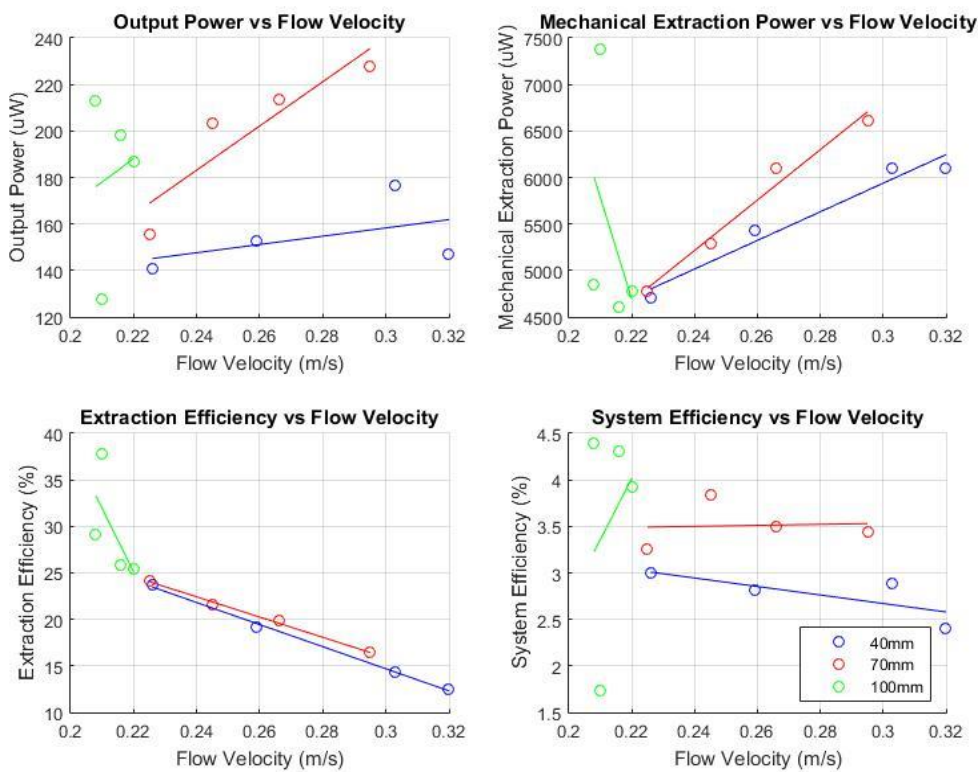

Fig. A.6. Power and efficiency for shedder distance 25 mm

Submergence = 40mm

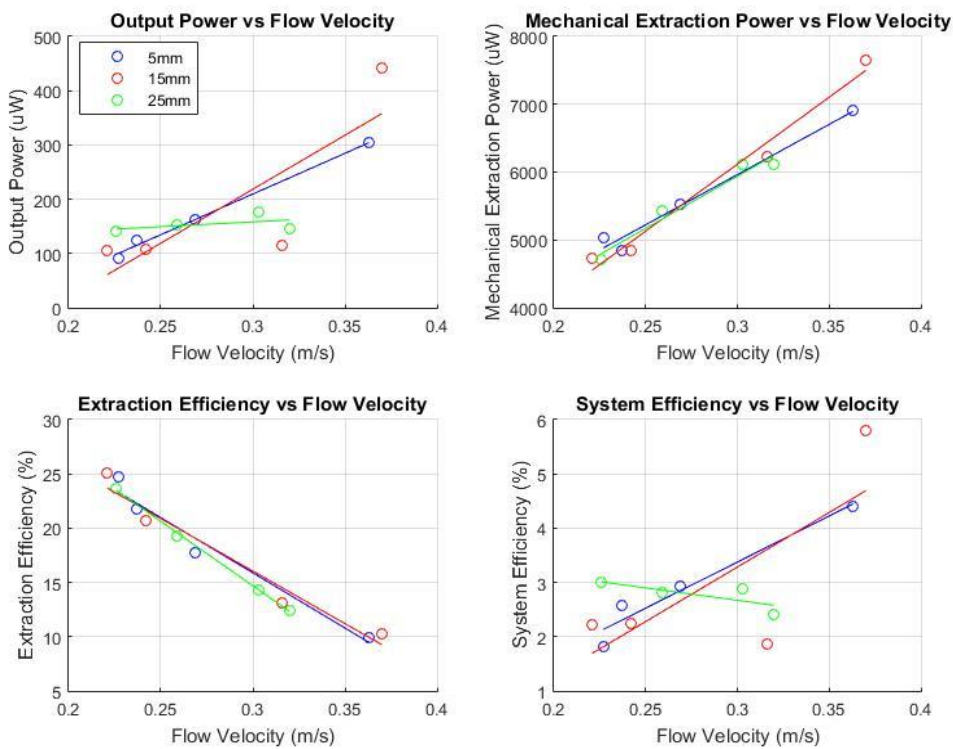

Fig. A.7. Power and efficiency for submergence 40 mm

Submergence = 70mm

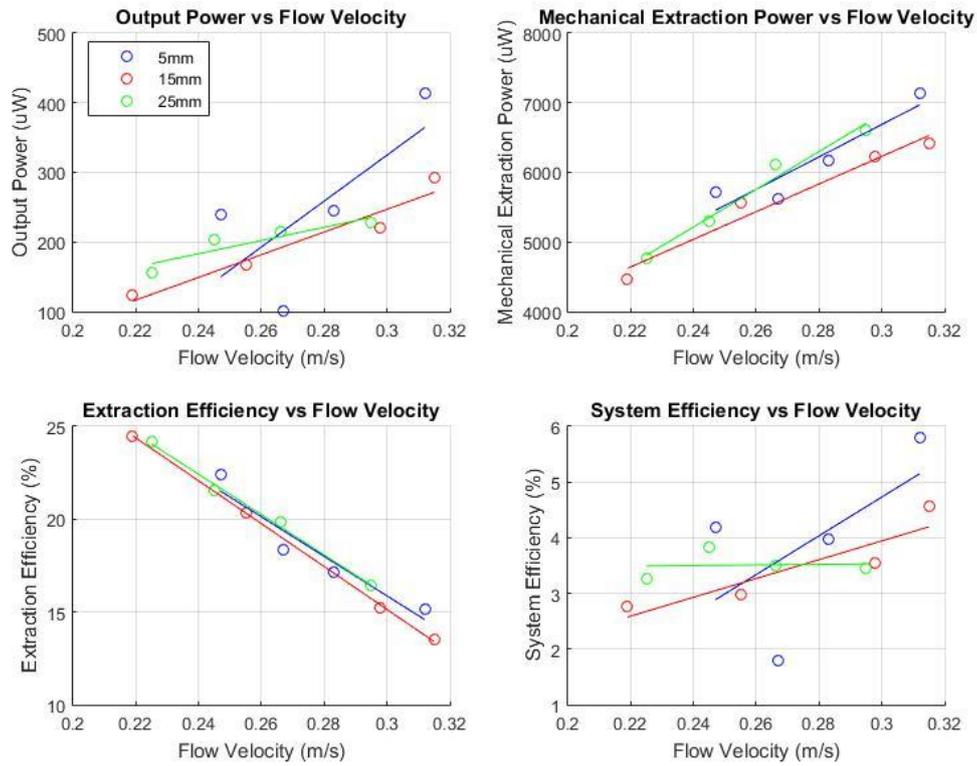

Fig. A.8. Power and efficiency for submergence 70 mm

Submergence = 100mm

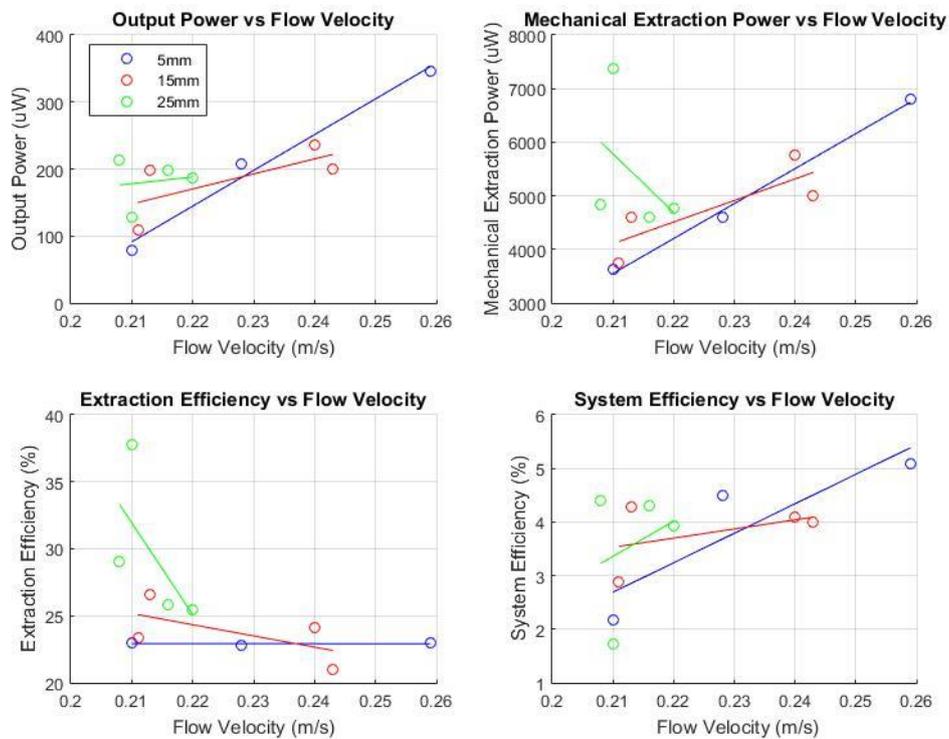

Fig. A.9. Power and efficiency for submergence 100 mm

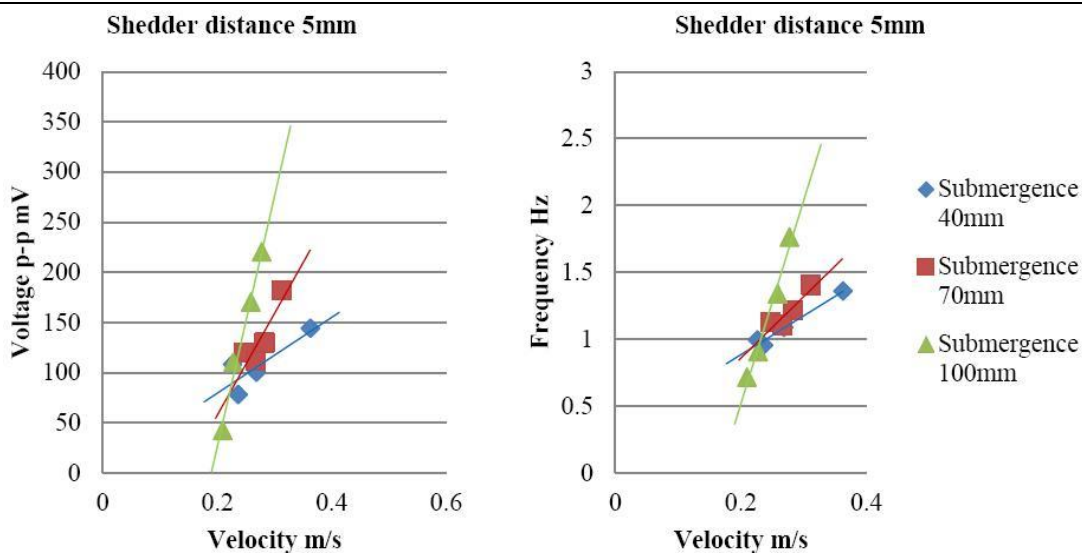

Fig. A.10. Voltage and frequency output at shedder distance 5 mm

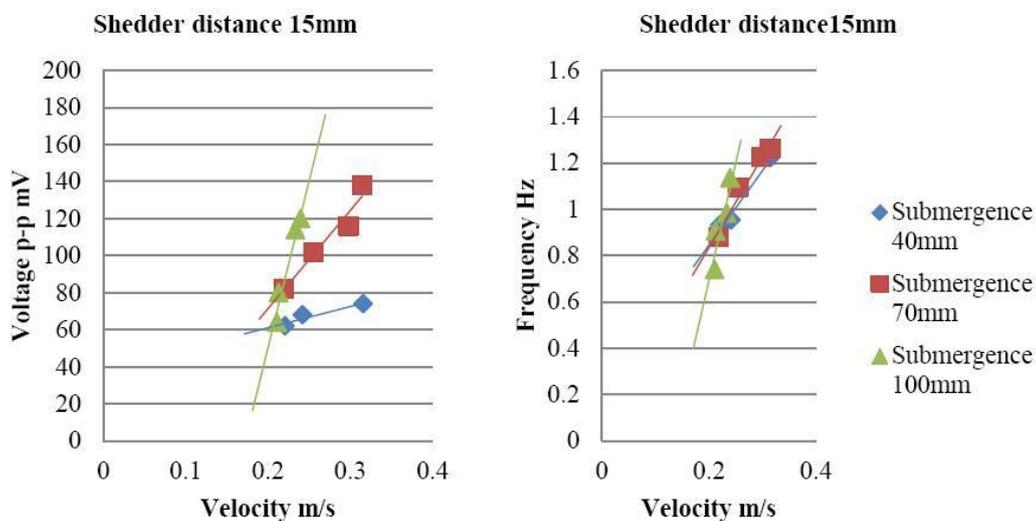

Fig. A.11. Voltage and frequency output at shedder distance 15 mm

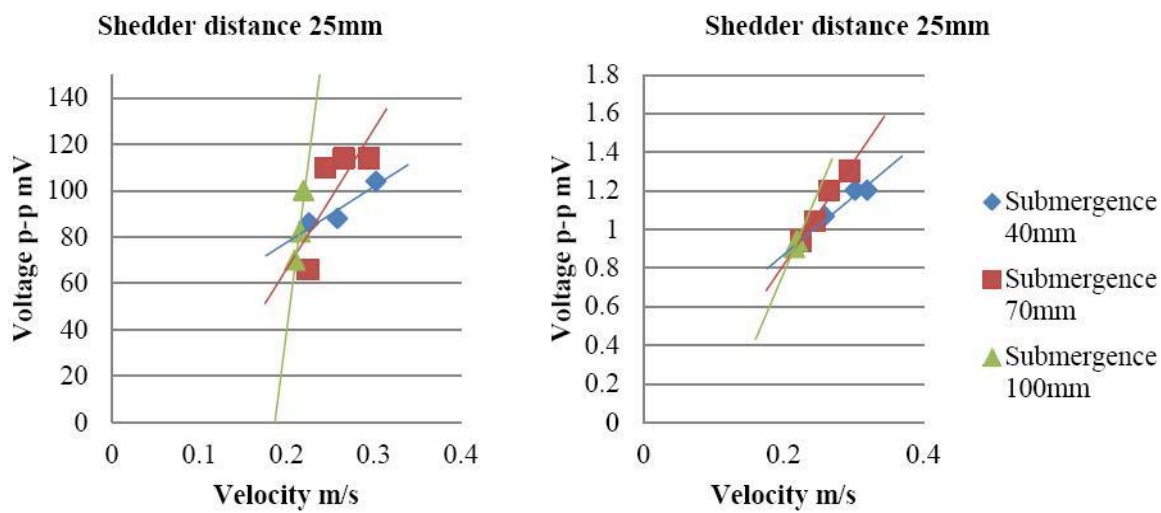

Fig. A.12. Voltage and frequency output at shedder distance 25 mm

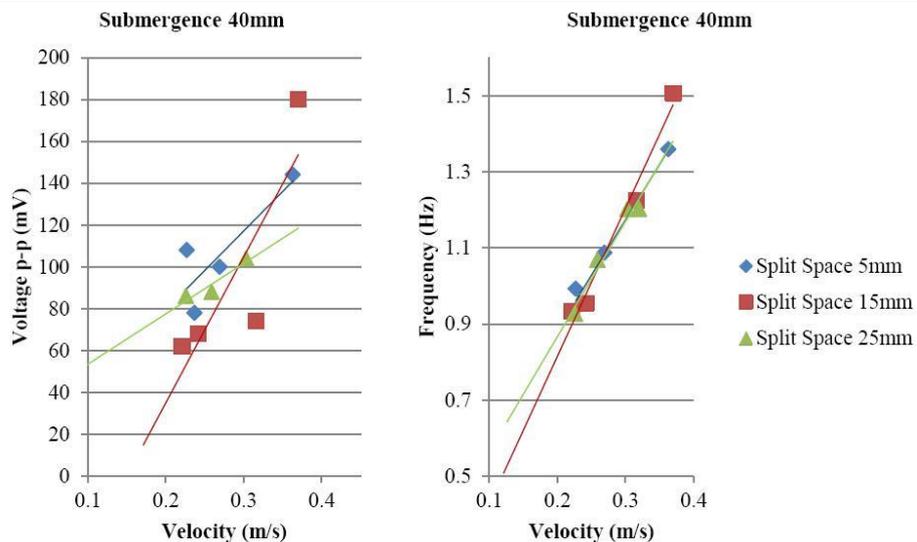

Fig. A.13. Voltage and frequency output for submergence 40 mm

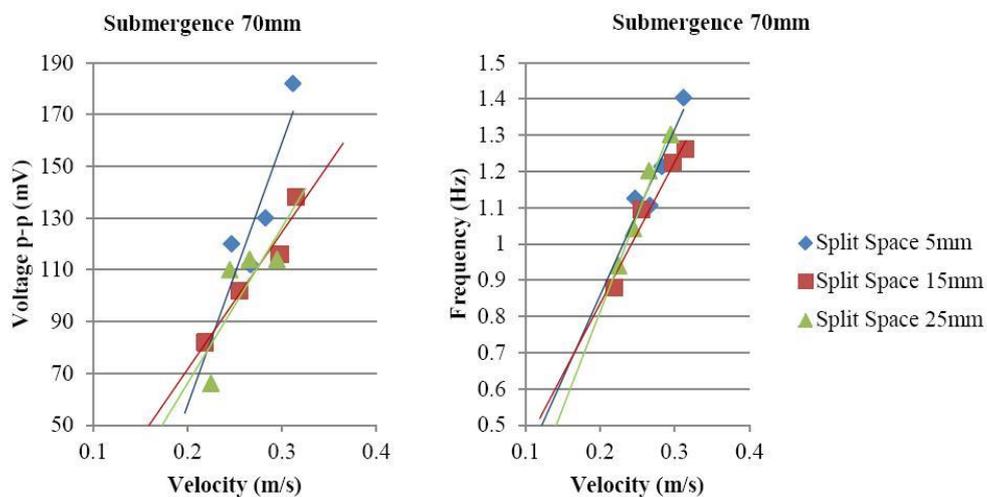

Fig. A.14. Voltage and frequency output for submergence 70 mm

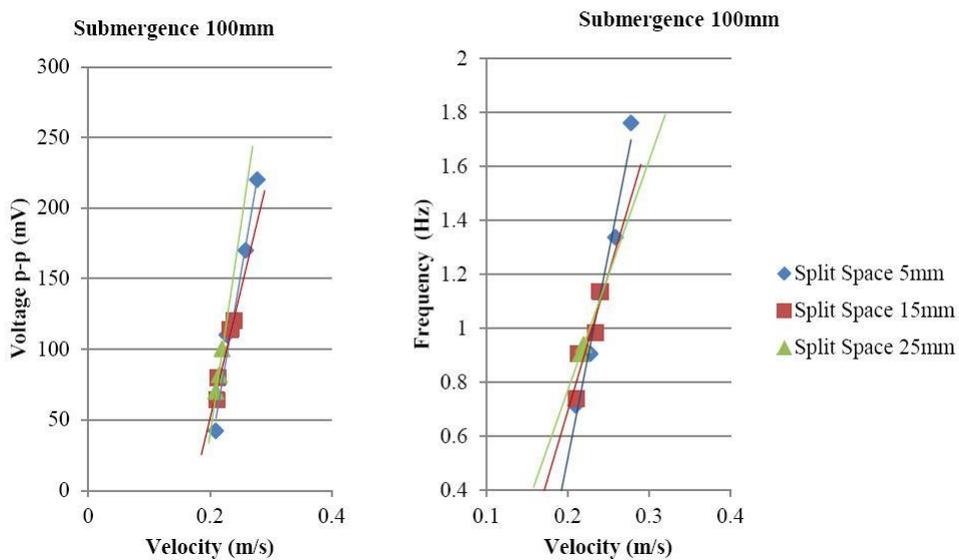

Fig. A.15. Voltage and frequency output for submergence 100 mm

**APPENDIX B**

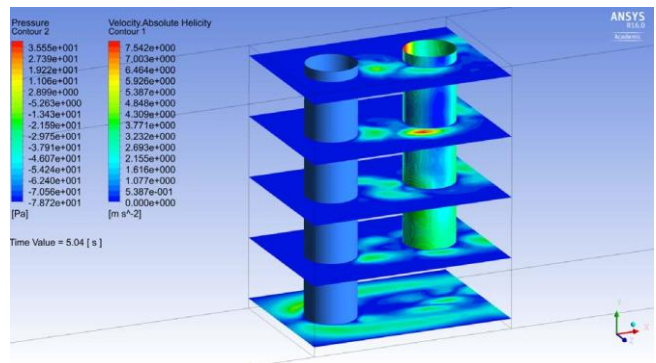

(a)

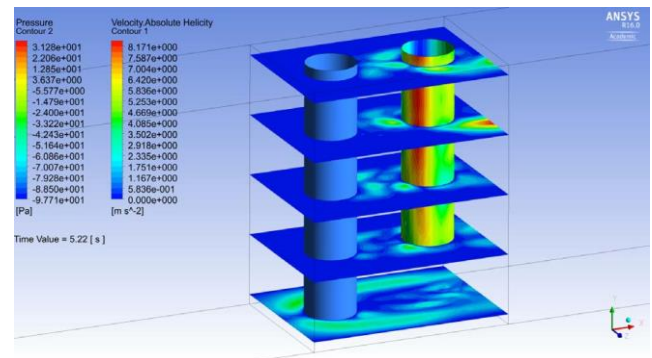

(b)

Fig. B.1. Pressure distribution for bluff body and absolute velocity helicity for planes of fluid at time (a) 5.04 s and (b) 5.22 s.

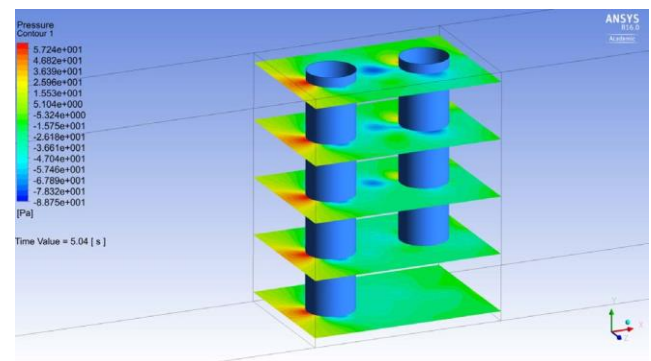

(a)

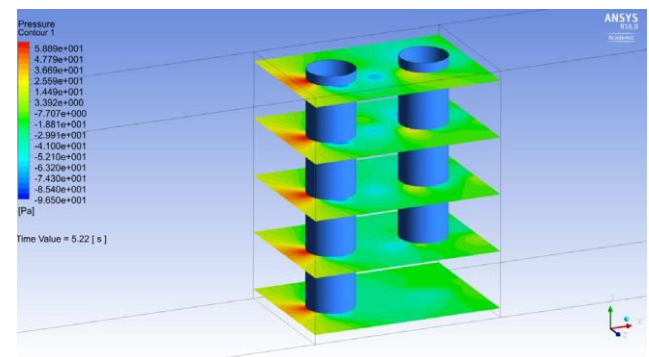

(b)

Fig. B.2. Pressure contours for planes of fluid at time (a) 5.04 s and (b) 5.22 s